\documentclass[aps,prl,twocolumn]{revtex4}  
\usepackage{graphicx}  
\usepackage{dcolumn}   
\usepackage{bm}        
\usepackage{amssymb}   
\usepackage{color} %
\RequirePackage{xspace}
\usepackage{relsize}
\usepackage{colortbl}
\usepackage{amsmath} 
\usepackage{ifthen} 
\usepackage{subfigure}

\usepackage{ifthen} 
\graphicspath{{ps}}
\newboolean{pdflatex}
\setboolean{pdflatex}{true} 
\usepackage{rotating} 

\newboolean{articletitles}
\setboolean{articletitles}{true} 

\newboolean{uprightparticles}
\setboolean{uprightparticles}{false} 

\usepackage{amssymb}
\usepackage{amsfonts}
\usepackage{upgreek} 

\usepackage{hyperref}    
\usepackage[all]{hypcap} 
\begin{document}



\def\lhcb {LHCb\xspace}
\def\ux85 {UX85\xspace}
\def\cern {CERN\xspace}
\def\lhc {LHC\xspace}
\def\atlas {ATLAS\xspace}
\def\cms {CMS\xspace}
\def\babar  {BaBar\xspace}
\def\belle  {Belle\xspace}
\def\aleph  {ALEPH\xspace}
\def\delphi {DELPHI\xspace}
\def\opal   {OPAL\xspace}
\def\lthree {L3\xspace}
\def\lep    {LEP\xspace}
\def\cdf    {CDF\xspace}
\def\dzero  {D\O\xspace}
\def\sld    {SLD\xspace}
\def\cleo   {CLEO\xspace}
\def\uaone  {UA1\xspace}
\def\uatwo  {UA2\xspace}
\def\tevatron {TEVATRON\xspace}


\def\pu     {PU\xspace}
\def\velo   {VELO\xspace}
\def\rich   {RICH\xspace}
\def\richone {RICH1\xspace}
\def\richtwo {RICH2\xspace}
\def\ttracker {TT\xspace}
\def\intr   {IT\xspace}
\def\st     {ST\xspace}
\def\ot     {OT\xspace}
\def\Tone   {T1\xspace}
\def\Ttwo   {T2\xspace}
\def\Tthree {T3\xspace}
\def\Mone   {M1\xspace}
\def\Mtwo   {M2\xspace}
\def\Mthree {M3\xspace}
\def\Mfour  {M4\xspace}
\def\Mfive  {M5\xspace}
\def\ecal   {ECAL\xspace}
\def\spd    {SPD\xspace}
\def\presh  {PS\xspace}
\def\hcal   {HCAL\xspace}
\def\bcm    {BCM\xspace}

\def\ode    {ODE\xspace}
\def\daq    {DAQ\xspace}
\def\tfc    {TFC\xspace}
\def\ecs    {ECS\xspace}
\def\lone   {L0\xspace}
\def\hlt    {HLT\xspace}
\def\hltone {HLT1\xspace}
\def\hlttwo {HLT2\xspace}


\ifthenelse{\boolean{uprightparticles}}%
{\def\Palpha      {\ensuremath{\upalpha}\xspace}
 \def\Pbeta       {\ensuremath{\upbeta}\xspace}
 \def\Pgamma      {\ensuremath{\upgamma}\xspace}                 
 \def\Pdelta      {\ensuremath{\updelta}\xspace}                 
 \def\Pepsilon    {\ensuremath{\upepsilon}\xspace}                 
 \def\Pvarepsilon {\ensuremath{\upvarepsilon}\xspace}                 
 \def\Pzeta       {\ensuremath{\upzeta}\xspace}                 
 \def\Peta        {\ensuremath{\upeta}\xspace}                 
 \def\Ptheta      {\ensuremath{\uptheta}\xspace}                 
 \def\Pvartheta   {\ensuremath{\upvartheta}\xspace}                 
 \def\Piota       {\ensuremath{\upiota}\xspace}                 
 \def\Pkappa      {\ensuremath{\upkappa}\xspace}                 
 \def\Plambda     {\ensuremath{\uplambda}\xspace}                 
 \def\Pmu         {\ensuremath{\upmu}\xspace}                 
 \def\Pnu         {\ensuremath{\upnu}\xspace}                 
 \def\Pxi         {\ensuremath{\upxi}\xspace}                 
 \def\Ppi         {\ensuremath{\uppi}\xspace}                 
 \def\Pvarpi      {\ensuremath{\upvarpi}\xspace}                 
 \def\Prho        {\ensuremath{\uprho}\xspace}                 
 \def\Pvarrho     {\ensuremath{\upvarrho}\xspace}                 
 \def\Ptau        {\ensuremath{\uptau}\xspace}                 
 \def\Pupsilon    {\ensuremath{\upupsilon}\xspace}                 
 \def\Pphi        {\ensuremath{\upphi}\xspace}                 
 \def\Pvarphi     {\ensuremath{\upvarphi}\xspace}                 
 \def\Pchi        {\ensuremath{\upchi}\xspace}                 
 \def\Ppsi        {\ensuremath{\uppsi}\xspace}                 
 \def\Pomega      {\ensuremath{\upomega}\xspace}                 

 \def\PDelta      {\ensuremath{\Delta}\xspace}                 
 \def\PXi      {\ensuremath{\Xi}\xspace}                 
 \def\PLambda      {\ensuremath{\Lambda}\xspace}                 
 \def\PSigma      {\ensuremath{\Sigma}\xspace}                 
 \def\POmega      {\ensuremath{\Omega}\xspace}                 
 \def\PUpsilon      {\ensuremath{\Upsilon}\xspace}                 
 

 \def\PA      {\ensuremath{\mathrm{A}}\xspace}                 
 \def\PB      {\ensuremath{\mathrm{B}}\xspace}                 
 \def\PC      {\ensuremath{\mathrm{C}}\xspace}                 
 \def\PD      {\ensuremath{\mathrm{D}}\xspace}                 
 \def\PE      {\ensuremath{\mathrm{E}}\xspace}                 
 \def\PF      {\ensuremath{\mathrm{F}}\xspace}                 
 \def\PG      {\ensuremath{\mathrm{G}}\xspace}                 
 \def\PH      {\ensuremath{\mathrm{H}}\xspace}                 
 \def\PI      {\ensuremath{\mathrm{I}}\xspace}                 
 \def\PJ      {\ensuremath{\mathrm{J}}\xspace}                 
 \def\PK      {\ensuremath{\mathrm{K}}\xspace}                 
 \def\PL      {\ensuremath{\mathrm{L}}\xspace}                 
 \def\PM      {\ensuremath{\mathrm{M}}\xspace}                 
 \def\PN      {\ensuremath{\mathrm{N}}\xspace}                 
 \def\PO      {\ensuremath{\mathrm{O}}\xspace}                 
 \def\PP      {\ensuremath{\mathrm{P}}\xspace}                 
 \def\PQ      {\ensuremath{\mathrm{Q}}\xspace}                 
 \def\PR      {\ensuremath{\mathrm{R}}\xspace}                 
 \def\PS      {\ensuremath{\mathrm{S}}\xspace}                 
 \def\PT      {\ensuremath{\mathrm{T}}\xspace}                 
 \def\PU      {\ensuremath{\mathrm{U}}\xspace}                 
 \def\PV      {\ensuremath{\mathrm{V}}\xspace}                 
 \def\PW      {\ensuremath{\mathrm{W}}\xspace}                 
 \def\PX      {\ensuremath{\mathrm{X}}\xspace}                 
 \def\PY      {\ensuremath{\mathrm{Y}}\xspace}                 
 \def\PZ      {\ensuremath{\mathrm{Z}}\xspace}                 
 \def\Pa      {\ensuremath{\mathrm{a}}\xspace}                 
 \def\Pb      {\ensuremath{\mathrm{b}}\xspace}                 
 \def\Pc      {\ensuremath{\mathrm{c}}\xspace}                 
 \def\Pd      {\ensuremath{\mathrm{d}}\xspace}                 
 \def\Pe      {\ensuremath{\mathrm{e}}\xspace}                 
 \def\Pf      {\ensuremath{\mathrm{f}}\xspace}                 
 \def\Pg      {\ensuremath{\mathrm{g}}\xspace}                 
 \def\Ph      {\ensuremath{\mathrm{h}}\xspace}                 
 \def\Pi      {\ensuremath{\mathrm{i}}\xspace}                 
 \def\Pj      {\ensuremath{\mathrm{j}}\xspace}                 
 \def\Pk      {\ensuremath{\mathrm{k}}\xspace}                 
 \def\Pl      {\ensuremath{\mathrm{l}}\xspace}                 
 \def\Pm      {\ensuremath{\mathrm{m}}\xspace}                 
 \def\Pn      {\ensuremath{\mathrm{n}}\xspace}                 
 \def\Po      {\ensuremath{\mathrm{o}}\xspace}                 
 \def\Pp      {\ensuremath{\mathrm{p}}\xspace}                 
 \def\Pq      {\ensuremath{\mathrm{q}}\xspace}                 
 \def\Pr      {\ensuremath{\mathrm{r}}\xspace}                 
 \def\Ps      {\ensuremath{\mathrm{s}}\xspace}                 
 \def\Pt      {\ensuremath{\mathrm{t}}\xspace}                 
 \def\Pu      {\ensuremath{\mathrm{u}}\xspace}                 
 \def\Pv      {\ensuremath{\mathrm{v}}\xspace}                 
 \def\Pw      {\ensuremath{\mathrm{w}}\xspace}                 
 \def\Px      {\ensuremath{\mathrm{x}}\xspace}                 
 \def\Py      {\ensuremath{\mathrm{y}}\xspace}                 
 \def\Pz      {\ensuremath{\mathrm{z}}\xspace}                 
}
{\def\Palpha      {\ensuremath{\alpha}\xspace}
 \def\Pbeta       {\ensuremath{\beta}\xspace}
 \def\Pgamma      {\ensuremath{\gamma}\xspace}                 
 \def\Pdelta      {\ensuremath{\delta}\xspace}                 
 \def\Pepsilon    {\ensuremath{\epsilon}\xspace}                 
 \def\Pvarepsilon {\ensuremath{\varepsilon}\xspace}                 
 \def\Pzeta       {\ensuremath{\zeta}\xspace}                 
 \def\Peta        {\ensuremath{\eta}\xspace}                 
 \def\Ptheta      {\ensuremath{\theta}\xspace}                 
 \def\Pvartheta   {\ensuremath{\vartheta}\xspace}                 
 \def\Piota       {\ensuremath{\iota}\xspace}                 
 \def\Pkappa      {\ensuremath{\kappa}\xspace}                 
 \def\Plambda     {\ensuremath{\lambda}\xspace}                 
 \def\Pmu         {\ensuremath{\mu}\xspace}                 
 \def\Pnu         {\ensuremath{\nu}\xspace}                 
 \def\Pxi         {\ensuremath{\xi}\xspace}                 
 \def\Ppi         {\ensuremath{\pi}\xspace}                 
 \def\Pvarpi      {\ensuremath{\varpi}\xspace}                 
 \def\Prho        {\ensuremath{\rho}\xspace}                 
 \def\Pvarrho     {\ensuremath{\varrho}\xspace}                 
 \def\Ptau        {\ensuremath{\tau}\xspace}                 
 \def\Pupsilon    {\ensuremath{\upsilon}\xspace}                 
 \def\Pphi        {\ensuremath{\phi}\xspace}                 
 \def\Pvarphi     {\ensuremath{\varphi}\xspace}                 
 \def\Pchi        {\ensuremath{\chi}\xspace}                 
 \def\Ppsi        {\ensuremath{\psi}\xspace}                 
 \def\Pomega      {\ensuremath{\omega}\xspace}                 
 \mathchardef\PDelta="7101
 \mathchardef\PXi="7104
 \mathchardef\PLambda="7103
 \mathchardef\PSigma="7106
 \mathchardef\POmega="710A
 \mathchardef\PUpsilon="7107
 \def\PA      {\ensuremath{A}\xspace}                 
 \def\PB      {\ensuremath{B}\xspace}                 
 \def\PC      {\ensuremath{C}\xspace}                 
 \def\PD      {\ensuremath{D}\xspace}                 
 \def\PE      {\ensuremath{E}\xspace}                 
 \def\PF      {\ensuremath{F}\xspace}                 
 \def\PG      {\ensuremath{G}\xspace}                 
 \def\PH      {\ensuremath{H}\xspace}                 
 \def\PI      {\ensuremath{I}\xspace}                 
 \def\PJ      {\ensuremath{J}\xspace}                 
 \def\PK      {\ensuremath{K}\xspace}                 
 \def\PL      {\ensuremath{L}\xspace}                 
 \def\PM      {\ensuremath{M}\xspace}                 
 \def\PN      {\ensuremath{N}\xspace}                 
 \def\PO      {\ensuremath{O}\xspace}                 
 \def\PP      {\ensuremath{P}\xspace}                 
 \def\PQ      {\ensuremath{Q}\xspace}                 
 \def\PR      {\ensuremath{R}\xspace}                 
 \def\PS      {\ensuremath{S}\xspace}                 
 \def\PT      {\ensuremath{T}\xspace}                 
 \def\PU      {\ensuremath{U}\xspace}                 
 \def\PV      {\ensuremath{V}\xspace}                 
 \def\PW      {\ensuremath{W}\xspace}                 
 \def\PX      {\ensuremath{X}\xspace}                 
 \def\PY      {\ensuremath{Y}\xspace}                 
 \def\PZ      {\ensuremath{Z}\xspace}                 
 \def\Pa      {\ensuremath{a}\xspace}                 
 \def\Pb      {\ensuremath{b}\xspace}                 
 \def\Pc      {\ensuremath{c}\xspace}                 
 \def\Pd      {\ensuremath{d}\xspace}                 
 \def\Pe      {\ensuremath{e}\xspace}                 
 \def\Pf      {\ensuremath{f}\xspace}                 
 \def\Pg      {\ensuremath{g}\xspace}                 
 \def\Ph      {\ensuremath{h}\xspace}                 
 \def\Pi      {\ensuremath{i}\xspace}                 
 \def\Pj      {\ensuremath{j}\xspace}                 
 \def\Pk      {\ensuremath{k}\xspace}                 
 \def\Pl      {\ensuremath{l}\xspace}                 
 \def\Pm      {\ensuremath{m}\xspace}                 
 \def\Pn      {\ensuremath{n}\xspace}                 
 \def\Po      {\ensuremath{o}\xspace}                 
 \def\Pp      {\ensuremath{p}\xspace}                 
 \def\Pq      {\ensuremath{q}\xspace}                 
 \def\Pr      {\ensuremath{r}\xspace}                 
 \def\Ps      {\ensuremath{s}\xspace}                 
 \def\Pt      {\ensuremath{t}\xspace}                 
 \def\Pu      {\ensuremath{u}\xspace}                 
 \def\Pv      {\ensuremath{v}\xspace}                 
 \def\Pw      {\ensuremath{w}\xspace}                 
 \def\Px      {\ensuremath{x}\xspace}                 
 \def\Py      {\ensuremath{y}\xspace}                 
 \def\Pz      {\ensuremath{z}\xspace}                 
}



\let\emi\en
\def\electron   {\ensuremath{\Pe}\xspace}
\def\en         {\ensuremath{\Pe^-}\xspace}   
\def\ep         {\ensuremath{\Pe^+}\xspace}
\def\epm        {\ensuremath{\Pe^\pm}\xspace} 
\def\epem       {\ensuremath{\Pe^+\Pe^-}\xspace}
\def\ee         {\ensuremath{\Pe^-\Pe^-}\xspace}

\def\mmu        {\ensuremath{\Pmu}\xspace}
\def\mup        {\ensuremath{\Pmu^+}\xspace}
\def\mun        {\ensuremath{\Pmu^-}\xspace} 
\def\mumu       {\ensuremath{\Pmu^+\Pmu^-}\xspace}
\def\mtau       {\ensuremath{\Ptau}\xspace}

\def\taup       {\ensuremath{\Ptau^+}\xspace}
\def\taum       {\ensuremath{\Ptau^-}\xspace}
\def\tautau     {\ensuremath{\Ptau^+\Ptau^-}\xspace}

\def\ellm       {\ensuremath{\ell^-}\xspace}
\def\ellp       {\ensuremath{\ell^+}\xspace}
\def\ellell     {\ensuremath{\ell^+ \ell^-}\xspace}

\def\neu        {\ensuremath{\Pnu}\xspace}
\def\neub       {\ensuremath{\overline{\Pnu}}\xspace}
\def\nuenueb    {\ensuremath{\neu\neub}\xspace}
\def\neue       {\ensuremath{\neu_e}\xspace}
\def\neueb      {\ensuremath{\neub_e}\xspace}
\def\neueneueb  {\ensuremath{\neue\neueb}\xspace}
\def\neum       {\ensuremath{\neu_\mu}\xspace}
\def\neumb      {\ensuremath{\neub_\mu}\xspace}
\def\neumneumb  {\ensuremath{\neum\neumb}\xspace}
\def\neut       {\ensuremath{\neu_\tau}\xspace}
\def\neutb      {\ensuremath{\neub_\tau}\xspace}
\def\neutneutb  {\ensuremath{\neut\neutb}\xspace}
\def\neul       {\ensuremath{\neu_\ell}\xspace}
\def\neulb      {\ensuremath{\neub_\ell}\xspace}
\def\neulneulb  {\ensuremath{\neul\neulb}\xspace}


\def\g      {\ensuremath{\Pgamma}\xspace}
\def\H      {\ensuremath{\PH^0}\xspace}
\def\Hp     {\ensuremath{\PH^+}\xspace}
\def\Hm     {\ensuremath{\PH^-}\xspace}
\def\Hpm    {\ensuremath{\PH^\pm}\xspace}
\def\W      {\ensuremath{\PW}\xspace}
\def\Wp     {\ensuremath{\PW^+}\xspace}
\def\Wm     {\ensuremath{\PW^-}\xspace}
\def\Wpm    {\ensuremath{\PW^\pm}\xspace}
\def\Z      {\ensuremath{\PZ^0}\xspace}


\def\quark     {\ensuremath{\Pq}\xspace}
\def\quarkbar  {\ensuremath{\overline \quark}\xspace}
\def\qqbar     {\ensuremath{\quark\quarkbar}\xspace}
\def\uquark    {\ensuremath{\Pu}\xspace}
\def\uquarkbar {\ensuremath{\overline \uquark}\xspace}
\def\uubar     {\ensuremath{\uquark\uquarkbar}\xspace}
\def\dquark    {\ensuremath{\Pd}\xspace}
\def\dquarkbar {\ensuremath{\overline \dquark}\xspace}
\def\ddbar     {\ensuremath{\dquark\dquarkbar}\xspace}
\def\squark    {\ensuremath{\Ps}\xspace}
\def\squarkbar {\ensuremath{\overline \squark}\xspace}
\def\ssbar     {\ensuremath{\squark\squarkbar}\xspace}
\def\cquark    {\ensuremath{\Pc}\xspace}
\def\cquarkbar {\ensuremath{\overline \cquark}\xspace}
\def\ccbar     {\ensuremath{\cquark\cquarkbar}\xspace}
\def\bquark    {\ensuremath{\Pb}\xspace}
\def\bquarkbar {\ensuremath{\overline \bquark}\xspace}
\def\bbbar     {\ensuremath{\bquark\bquarkbar}\xspace}
\def\tquark    {\ensuremath{\Pt}\xspace}
\def\tquarkbar {\ensuremath{\overline \tquark}\xspace}
\def\ttbar     {\ensuremath{\tquark\tquarkbar}\xspace}


\def\pion  {\ensuremath{\Ppi}\xspace}
\def\piz   {\ensuremath{\pion^0}\xspace}
\def\pizs  {\ensuremath{\pion^0\mbox\,\rm{s}}\xspace}
\def\ppz   {\ensuremath{\pion^0\pion^0}\xspace}
\def\pip   {\ensuremath{\pion^+}\xspace}
\def\pim   {\ensuremath{\pion^-}\xspace}
\def\pipi  {\ensuremath{\pion^+\pion^-}\xspace}
\def\pipm  {\ensuremath{\pion^\pm}\xspace}
\def\pimp  {\ensuremath{\pion^\mp}\xspace}

\def\kaon  {\ensuremath{\PK}\xspace}
  \def\Kbar  {\kern 0.2em\overline{\kern -0.2em \PK}{}\xspace}
\def\Kb    {\ensuremath{\Kbar}\xspace}
\def\Kz    {\ensuremath{\kaon^0}\xspace}
\def\Kzb   {\ensuremath{\Kbar^0}\xspace}
\def\KzKzb {\ensuremath{\Kz \kern -0.16em \Kzb}\xspace}
\def\Kp    {\ensuremath{\kaon^+}\xspace}
\def\Km    {\ensuremath{\kaon^-}\xspace}
\def\Kpm   {\ensuremath{\kaon^\pm}\xspace}
\def\Kmp   {\ensuremath{\kaon^\mp}\xspace}
\def\KpKm  {\ensuremath{\Kp \kern -0.16em \Km}\xspace}
\def\KS    {\ensuremath{\kaon^0_{\rm\scriptscriptstyle S}}\xspace} 
\def\KSb    {\ensuremath{\Kbar^0_{\rm\scriptscriptstyle S}}\xspace} 
\def\KL    {\ensuremath{\kaon^0_{\rm\scriptscriptstyle L}}\xspace} 
\def\Kstarz  {\ensuremath{\kaon^{*0}}\xspace}
\def\Kstarzb {\ensuremath{\Kbar^{*0}}\xspace}
\def\Kstar   {\ensuremath{\kaon^*}\xspace}
\def\Kstarb  {\ensuremath{\Kbar^*}\xspace}
\def\Kstarp  {\ensuremath{\kaon^{*+}}\xspace}
\def\Kstarm  {\ensuremath{\kaon^{*-}}\xspace}
\def\Kstarpm {\ensuremath{\kaon^{*\pm}}\xspace}
\def\Kstarmp {\ensuremath{\kaon^{*\mp}}\xspace}

\newcommand{\etapr}{\ensuremath{\Peta^{\prime}}\xspace}


  \def\Dbar    {\kern 0.2em\overline{\kern -0.2em \PD}{}\xspace}
\def\D       {\ensuremath{\PD}\xspace}
\def\Db      {\ensuremath{\Dbar}\xspace}
\def\Dz      {\ensuremath{\D^0}\xspace}
\def\Dzb     {\ensuremath{\Dbar^0}\xspace}
\def\DzDzb   {\ensuremath{\Dz {\kern -0.16em \Dzb}}\xspace}
\def\Dp      {\ensuremath{\D^+}\xspace}
\def\Dm      {\ensuremath{\D^-}\xspace}
\def\Dpm     {\ensuremath{\D^\pm}\xspace}
\def\Dmp     {\ensuremath{\D^\mp}\xspace}
\def\DpDm    {\ensuremath{\Dp {\kern -0.16em \Dm}}\xspace}
\def\Dstar   {\ensuremath{\D^*}\xspace}
\def\Dstarb  {\ensuremath{\Dbar^*}\xspace}
\def\Dstarz  {\ensuremath{\D^{*0}}\xspace}
\def\Dstarzb {\ensuremath{\Dbar^{*0}}\xspace}
\def\Dstarp  {\ensuremath{\D^{*+}}\xspace}
\def\Dstarm  {\ensuremath{\D^{*-}}\xspace}
\def\Dstarpm {\ensuremath{\D^{*\pm}}\xspace}
\def\Dstarmp {\ensuremath{\D^{*\mp}}\xspace}
\def\Ds      {\ensuremath{\D^+_\squark}\xspace}
\def\Dsp     {\ensuremath{\D^+_\squark}\xspace}
\def\Dsm     {\ensuremath{\D^-_\squark}\xspace}
\def\Dspm    {\ensuremath{\D^{\pm}_\squark}\xspace}
\def\Dss     {\ensuremath{\D^{*+}_\squark}\xspace}
\def\Dssp    {\ensuremath{\D^{*+}_\squark}\xspace}
\def\Dssm    {\ensuremath{\D^{*-}_\squark}\xspace}
\def\Dsspm   {\ensuremath{\D^{*\pm}_\squark}\xspace}

\def\B       {\ensuremath{\PB}\xspace}
  \def\Bbar    {\kern 0.18em\overline{\kern -0.18em \PB}{}\xspace}
\def\Bb      {\ensuremath{\Bbar}\xspace}
\def\BBbar   {\ensuremath{\B\Bbar}\xspace} 
\def\Bz      {\ensuremath{\B^0}\xspace}
\def\Bzb     {\ensuremath{\Bbar^0}\xspace}
\def\Bu      {\ensuremath{\B^+}\xspace}
\def\Bub     {\ensuremath{\B^-}\xspace}
\def\Bp      {\ensuremath{\Bu}\xspace}
\def\Bm      {\ensuremath{\Bub}\xspace}
\def\Bpm     {\ensuremath{\B^\pm}\xspace}
\def\Bmp     {\ensuremath{\B^\mp}\xspace}
\def\Bd      {\ensuremath{\B^0}\xspace}
\def\Bs      {\ensuremath{\B^0_\squark}\xspace}
\def\Bsb     {\ensuremath{\Bbar^0_\squark}\xspace}
\def\Bstar   {\ensuremath{B_s^*}\xspace}
\def\Bstarb  {\ensuremath{\Bb_s^*}\xspace}
\def\Bdb     {\ensuremath{\Bbar^0}\xspace}
\def\Bc      {\ensuremath{\B_\cquark^+}\xspace}
\def\Bcp     {\ensuremath{\B_\cquark^+}\xspace}
\def\Bcm     {\ensuremath{\B_\cquark^-}\xspace}
\def\Bcpm    {\ensuremath{\B_\cquark^\pm}\xspace}


\def\jpsi     {\ensuremath{{\PJ\mskip -3mu/\mskip -2mu\Ppsi\mskip 2mu}}\xspace}
\def\psitwos  {\ensuremath{\Ppsi{(2S)}}\xspace}
\def\psiprpr  {\ensuremath{\Ppsi(3770)}\xspace}
\def\etac     {\ensuremath{\Peta_\cquark}\xspace}
\def\chiczero {\ensuremath{\Pchi_{\cquark 0}}\xspace}
\def\chicone  {\ensuremath{\Pchi_{\cquark 1}}\xspace}
\def\chictwo  {\ensuremath{\Pchi_{\cquark 2}}\xspace}
  \def\Y#1S{\ensuremath{\PUpsilon{(#1S)}}\xspace}
\def\OneS  {\Y1S}
\def\TwoS  {\Y2S}
\def\ThreeS{\Y3S}
\def\FourS {\Y4S}
\def\FiveS {\Y5S}

\def\chic  {\ensuremath{\Pchi_{c}}\xspace}


\def\proton      {\ensuremath{\Pp}\xspace}
\def\antiproton  {\ensuremath{\overline \proton}\xspace}
\def\neutron     {\ensuremath{\Pn}\xspace}
\def\antineutron {\ensuremath{\overline \neutron}\xspace}

\def\Deltares {\ensuremath{\PDelta}\xspace}
\def\Deltaresbar{\ensuremath{\overline \Deltares}\xspace}
\def\Xires {\ensuremath{\PXi}\xspace}
\def\Xiresbar{\ensuremath{\overline \Xires}\xspace}
\def\L {\ensuremath{\PLambda}\xspace}
\def\Lbar {\ensuremath{\kern 0.1em\overline{\kern -0.1em\Lambda\kern -0.05em}\kern 0.05em{}}\xspace}
\def\Lambdares {\ensuremath{\PLambda}\xspace}
\def\Lambdaresbar{\ensuremath{\Lbar}\xspace}
\def\Sigmares {\ensuremath{\PSigma}\xspace}
\def\Sigmaresbar{\ensuremath{\overline \Sigmares}\xspace}
\def\Omegares {\ensuremath{\POmega}\xspace}
\def\Omegaresbar{\ensuremath{\overline \Omegares}\xspace}


\def\Lb      {\ensuremath{\L^0_\bquark}\xspace}
\def\Lbbar   {\ensuremath{\Lbar^0_\bquark}\xspace}
\def\Lc      {\ensuremath{\L^+_\cquark}\xspace}
\def\Lcbar   {\ensuremath{\Lbar^-_\cquark}\xspace}


\def\BF         {{\ensuremath{\cal B}\xspace}}
\def\BRvis      {{\ensuremath{\BR_{\rm{vis}}}}}
\def\BR         {\BF}
\newcommand{\decay}[2]{\ensuremath{#1\!\to #2}\xspace}         
\def\ra                 {\ensuremath{\rightarrow}\xspace}
\def\to                 {\ensuremath{\rightarrow}\xspace}

\newcommand{\tauBs}{\ensuremath{\tau_{\Bs}}\xspace}
\newcommand{\tauBd}{\ensuremath{\tau_{\Bd}}\xspace}
\newcommand{\tauBz}{\ensuremath{\tau_{\Bz}}\xspace}
\newcommand{\tauBu}{\ensuremath{\tau_{\Bp}}\xspace}
\newcommand{\tauDp}{\ensuremath{\tau_{\Dp}}\xspace}
\newcommand{\tauDz}{\ensuremath{\tau_{\Dz}}\xspace}
\newcommand{\tauL}{\ensuremath{\tau_{\rm L}}\xspace}
\newcommand{\tauH}{\ensuremath{\tau_{\rm H}}\xspace}

\newcommand{\mBd}{\ensuremath{m_{\Bd}}\xspace}
\newcommand{\mBp}{\ensuremath{m_{\Bp}}\xspace}
\newcommand{\mBs}{\ensuremath{m_{\Bs}}\xspace}
\newcommand{\mBc}{\ensuremath{m_{\Bc}}\xspace}
\newcommand{\mLb}{\ensuremath{m_{\Lb}}\xspace}

\def\grpsuthree {\ensuremath{\mathrm{SU}(3)}\xspace}
\def\grpsutw    {\ensuremath{\mathrm{SU}(2)}\xspace}
\def\grpuone    {\ensuremath{\mathrm{U}(1)}\xspace}

\def\ssqtw {\ensuremath{\sin^{2}\!\theta_{\mathrm{W}}}\xspace}
\def\csqtw {\ensuremath{\cos^{2}\!\theta_{\mathrm{W}}}\xspace}
\def\stw   {\ensuremath{\sin\theta_{\mathrm{W}}}\xspace}
\def\ctw   {\ensuremath{\cos\theta_{\mathrm{W}}}\xspace}
\def\ssqtwef {\ensuremath{{\sin}^{2}\theta_{\mathrm{W}}^{\mathrm{eff}}}\xspace}
\def\csqtwef {\ensuremath{{\cos}^{2}\theta_{\mathrm{W}}^{\mathrm{eff}}}\xspace}
\def\stwef {\ensuremath{\sin\theta_{\mathrm{W}}^{\mathrm{eff}}}\xspace}
\def\ctwef {\ensuremath{\cos\theta_{\mathrm{W}}^{\mathrm{eff}}}\xspace}
\def\gv    {\ensuremath{g_{\mbox{\tiny V}}}\xspace}
\def\ga    {\ensuremath{g_{\mbox{\tiny A}}}\xspace}

\def\order   {\ensuremath{\mathcal{O}}\xspace}
\def\ordalph {\ensuremath{\mathcal{O}(\alpha)}\xspace}
\def\ordalsq {\ensuremath{\mathcal{O}(\alpha^{2})}\xspace}
\def\ordalcb {\ensuremath{\mathcal{O}(\alpha^{3})}\xspace}

\newcommand{\as}{\ensuremath{\alpha_{\scriptscriptstyle S}}\xspace}
\newcommand{\MSb}{\ensuremath{\overline{\mathrm{MS}}}\xspace}
\newcommand{\lqcd}{\ensuremath{\Lambda_{\mathrm{QCD}}}\xspace}
\def\qsq       {\ensuremath{q^2}\xspace}


\def\eps   {\ensuremath{\varepsilon}\xspace}
\def\epsK  {\ensuremath{\varepsilon_K}\xspace}
\def\epsB  {\ensuremath{\varepsilon_B}\xspace}
\def\epsp  {\ensuremath{\varepsilon^\prime_K}\xspace}

\def\CP                {\ensuremath{C\!P}\xspace}
\def\CPT               {\ensuremath{C\!PT}\xspace}

\def\rhobar {\ensuremath{\overline \rho}\xspace}
\def\etabar {\ensuremath{\overline \eta}\xspace}

\def\Vud  {\ensuremath{|V_{\uquark\dquark}|}\xspace}
\def\Vcd  {\ensuremath{|V_{\cquark\dquark}|}\xspace}
\def\Vtd  {\ensuremath{|V_{\tquark\dquark}|}\xspace}
\def\Vus  {\ensuremath{|V_{\uquark\squark}|}\xspace}
\def\Vcs  {\ensuremath{|V_{\cquark\squark}|}\xspace}
\def\Vts  {\ensuremath{|V_{\tquark\squark}|}\xspace}
\def\Vub  {\ensuremath{|V_{\uquark\bquark}|}\xspace}
\def\Vcb  {\ensuremath{|V_{\cquark\bquark}|}\xspace}
\def\Vtb  {\ensuremath{|V_{\tquark\bquark}|}\xspace}


\newcommand{\dm}{\ensuremath{\Delta m}\xspace}
\newcommand{\dms}{\ensuremath{\Delta m_{\squark}}\xspace}
\newcommand{\dmd}{\ensuremath{\Delta m_{\dquark}}\xspace}
\newcommand{\DG}{\ensuremath{\Delta\Gamma}\xspace}
\newcommand{\DGs}{\ensuremath{\Delta\Gamma_{\squark}}\xspace}
\newcommand{\DGd}{\ensuremath{\Delta\Gamma_{\dquark}}\xspace}
\newcommand{\Gs}{\ensuremath{\Gamma_{\squark}}\xspace}
\newcommand{\Gd}{\ensuremath{\Gamma_{\dquark}}\xspace}

\newcommand{\MBq}{\ensuremath{M_{\B_\quark}}\xspace}
\newcommand{\DGq}{\ensuremath{\Delta\Gamma_{\quark}}\xspace}
\newcommand{\Gq}{\ensuremath{\Gamma_{\quark}}\xspace}
\newcommand{\dmq}{\ensuremath{\Delta m_{\quark}}\xspace}
\newcommand{\GL}{\ensuremath{\Gamma_{\rm L}}\xspace}
\newcommand{\GH}{\ensuremath{\Gamma_{\rm H}}\xspace}

\newcommand{\DGsGs}{\ensuremath{\Delta\Gamma_{\squark}/\Gamma_{\squark}}\xspace}
\newcommand{\Delm}{\mbox{$\Delta m $}\xspace}
\newcommand{\ACP}{\ensuremath{{\cal A}^{\CP}}\xspace}
\newcommand{\Adir}{\ensuremath{{\cal A}^{\rm dir}}\xspace}
\newcommand{\Amix}{\ensuremath{{\cal A}^{\rm mix}}\xspace}
\newcommand{\ADelta}{\ensuremath{{\cal A}^\Delta}\xspace}
\newcommand{\phid}{\ensuremath{\phi_{\dquark}}\xspace}
\newcommand{\sinphid}{\ensuremath{\sin\!\phid}\xspace}
\newcommand{\phis}{\ensuremath{\phi_{\squark}}\xspace}
\newcommand{\betas}{\ensuremath{\beta_{\squark}}\xspace}
\newcommand{\sbetas}{\ensuremath{\sigma(\beta_{\squark})}\xspace}
\newcommand{\stbetas}{\ensuremath{\sigma(2\beta_{\squark})}\xspace}
\newcommand{\stphis}{\ensuremath{\sigma(\phi_{\squark})}\xspace}
\newcommand{\sinphis}{\ensuremath{\sin\!\phis}\xspace}

\newcommand{\edet}{{\ensuremath{\varepsilon_{\rm det}}}\xspace}
\newcommand{\erec}{{\ensuremath{\varepsilon_{\rm rec/det}}}\xspace}
\newcommand{\esel}{{\ensuremath{\varepsilon_{\rm sel/rec}}}\xspace}
\newcommand{\etrg}{{\ensuremath{\varepsilon_{\rm trg/sel}}}\xspace}
\newcommand{\etot}{{\ensuremath{\varepsilon_{\rm tot}}}\xspace}

\newcommand{\mistag}{\ensuremath{\omega}\xspace}
\newcommand{\wcomb}{\ensuremath{\omega^{\rm comb}}\xspace}
\newcommand{\etag}{{\ensuremath{\varepsilon_{\rm tag}}}\xspace}
\newcommand{\etagcomb}{{\ensuremath{\varepsilon_{\rm tag}^{\rm comb}}}\xspace}
\newcommand{\effeff}{\ensuremath{\varepsilon_{\rm eff}}\xspace}
\newcommand{\effeffcomb}{\ensuremath{\varepsilon_{\rm eff}^{\rm comb}}\xspace}
\newcommand{\efftag}{{\ensuremath{\etag(1-2\omega)^2}}\xspace}
\newcommand{\effD}{{\ensuremath{\etag D^2}}\xspace}

\newcommand{\etagprompt}{{\ensuremath{\varepsilon_{\rm tag}^{\rm Pr}}}\xspace}
\newcommand{\etagLL}{{\ensuremath{\varepsilon_{\rm tag}^{\rm LL}}}\xspace}


\def\BdToKstmm    {\decay{\Bd}{\Kstarz\mup\mun}}
\def\BdbToKstmm   {\decay{\Bdb}{\Kstarzb\mup\mun}}

\def\BsToJPsiPhi  {\decay{\Bs}{\jpsi\phi}}
\def\BdToJPsiKst  {\decay{\Bd}{\jpsi\Kstarz}}
\def\BdbToJPsiKst {\decay{\Bdb}{\jpsi\Kstarzb}}

\def\BsPhiGam     {\decay{\Bs}{\phi \g}}
\def\BdKstGam     {\decay{\Bd}{\Kstarz \g}}

\def\BTohh        {\decay{\B}{\Ph^+ \Ph'^-}}
\def\BdTopipi     {\decay{\Bd}{\pip\pim}}
\def\BdToKpi      {\decay{\Bd}{\Kp\pim}}
\def\BsToKK       {\decay{\Bs}{\Kp\Km}}
\def\BsTopiK      {\decay{\Bs}{\pip\Km}}

\def\BdKstee  {\decay{\Bd}{\Kstarz\epem}}
\def\BdbKstee {\decay{\Bdb}{\Kstarzb\epem}}
\def\bsll     {\decay{\bquark}{\squark \ell^+ \ell^-}}
\def\AFB      {\ensuremath{A_{\mathrm{FB}}}\xspace}
\def\FL       {\ensuremath{F_{\mathrm{L}}}\xspace}
\def\AT#1     {\ensuremath{A_{\mathrm{T}}^{#1}}\xspace}           
\def\btosgam  {\decay{\bquark}{\squark \g}}
\def\btodgam  {\decay{\bquark}{\dquark \g}}
\def\Bsmm     {\decay{\Bs}{\mup\mun}}
\def\Bdmm     {\decay{\Bd}{\mup\mun}}
\def\ctl       {\ensuremath{\cos{\theta_l}}\xspace}
\def\ctk       {\ensuremath{\cos{\theta_K}}\xspace}

\def\C#1      {\ensuremath{\mathcal{C}_{#1}}\xspace}                       
\def\Cp#1     {\ensuremath{\mathcal{C}_{#1}^{'}}\xspace}                    
\def\Ceff#1   {\ensuremath{\mathcal{C}_{#1}^{\mathrm{(eff)}}}\xspace}        
\def\Cpeff#1  {\ensuremath{\mathcal{C}_{#1}^{'\mathrm{(eff)}}}\xspace}       
\def\Ope#1    {\ensuremath{\mathcal{O}_{#1}}\xspace}                       
\def\Opep#1   {\ensuremath{\mathcal{O}_{#1}^{'}}\xspace}                    


\def\xprime     {\ensuremath{x^{\prime}}\xspace}
\def\yprime     {\ensuremath{y^{\prime}}\xspace}
\def\ycp        {\ensuremath{y_{\CP}}\xspace}
\def\agamma     {\ensuremath{A_{\Gamma}}\xspace}
\def\kpi        {\ensuremath{\PK\Ppi}\xspace}
\def\kk         {\ensuremath{\PK\PK}\xspace}
\def\dkpi       {\decay{\PD}{\PK\Ppi}}
\def\dkk        {\decay{\PD}{\PK\PK}}
\def\dkpicf     {\decay{\Dz}{\Km\pip}}

\newcommand{\bra}[1]{\ensuremath{\langle #1|}}             
\newcommand{\ket}[1]{\ensuremath{|#1\rangle}}              
\newcommand{\braket}[2]{\ensuremath{\langle #1|#2\rangle}} 

\newcommand{\unit}[1]{\ensuremath{\rm\,#1}\xspace}          

\newcommand{\tev}{\ensuremath{\mathrm{\,Te\kern -0.1em V}}\xspace}
\newcommand{\gev}{\ensuremath{\mathrm{\,Ge\kern -0.1em V}}\xspace}
\newcommand{\mev}{\ensuremath{\mathrm{\,Me\kern -0.1em V}}\xspace}
\newcommand{\kev}{\ensuremath{\mathrm{\,ke\kern -0.1em V}}\xspace}
\newcommand{\ev}{\ensuremath{\mathrm{\,e\kern -0.1em V}}\xspace}
\newcommand{\gevc}{\ensuremath{{\mathrm{\,Ge\kern -0.1em V\!/}c}}\xspace}
\newcommand{\mevc}{\ensuremath{{\mathrm{\,Me\kern -0.1em V\!/}c}}\xspace}
\newcommand{\gevcc}{\ensuremath{{\mathrm{\,Ge\kern -0.1em V\!/}c^2}}\xspace}
\newcommand{\gevgevcccc}{\ensuremath{{\mathrm{\,Ge\kern -0.1em V^2\!/}c^4}}\xspace}
\newcommand{\mevcc}{\ensuremath{{\mathrm{\,Me\kern -0.1em V\!/}c^2}}\xspace}

\def\km   {\ensuremath{\rm \,km}\xspace}
\def\m    {\ensuremath{\rm \,m}\xspace}
\def\cm   {\ensuremath{\rm \,cm}\xspace}
\def\cma  {\ensuremath{{\rm \,cm}^2}\xspace}
\def\mm   {\ensuremath{\rm \,mm}\xspace}
\def\mma  {\ensuremath{{\rm \,mm}^2}\xspace}
\def\mum  {\ensuremath{\,\upmu\rm m}\xspace}
\def\muma {\ensuremath{\,\upmu\rm m^2}\xspace}
\def\nm   {\ensuremath{\rm \,nm}\xspace}
\def\fm   {\ensuremath{\rm \,fm}\xspace}
\def\barn{\ensuremath{\rm \,b}\xspace}
\def\barnhyph{\ensuremath{\rm -b}\xspace}
\def\mbarn{\ensuremath{\rm \,mb}\xspace}
\def\mub{\ensuremath{\rm \,\upmu b}\xspace}
\def\mbarnhyph{\ensuremath{\rm -mb}\xspace}
\def\nb {\ensuremath{\rm \,nb}\xspace}
\def\invnb {\ensuremath{\mbox{\,nb}^{-1}}\xspace}
\def\pb {\ensuremath{\rm \,pb}\xspace}
\def\invpb {\ensuremath{\mbox{\,pb}^{-1}}\xspace}
\def\fb   {\ensuremath{\mbox{\,fb}}\xspace}
\def\invfb   {\ensuremath{\mbox{\,fb}^{-1}}\xspace}

\def\sec  {\ensuremath{\rm {\,s}}\xspace}
\def\ms   {\ensuremath{{\rm \,ms}}\xspace}
\def\mus  {\ensuremath{\,\upmu{\rm s}}\xspace}
\def\ns   {\ensuremath{{\rm \,ns}}\xspace}
\def\ps   {\ensuremath{{\rm \,ps}}\xspace}
\def\fs   {\ensuremath{\rm \,fs}\xspace}

\def\mhz  {\ensuremath{{\rm \,MHz}}\xspace}
\def\khz  {\ensuremath{{\rm \,kHz}}\xspace}
\def\hz   {\ensuremath{{\rm \,Hz}}\xspace}

\def\invps{\ensuremath{{\rm \,ps^{-1}}}\xspace}

\def\yr   {\ensuremath{\rm \,yr}\xspace}
\def\hr   {\ensuremath{\rm \,hr}\xspace}

\def\degc {\ensuremath{^\circ}{C}\xspace}
\def\degk {\ensuremath {\rm K}\xspace}

\def\Xrad {\ensuremath{X_0}\xspace}
\def\NIL{\ensuremath{\lambda_{int}}\xspace}
\def\mip {MIP\xspace}
\def\neutroneq {\ensuremath{\rm \,n_{eq}}\xspace}
\def\neqcmcm {\ensuremath{\rm \,n_{eq} / cm^2}\xspace}
\def\kRad {\ensuremath{\rm \,kRad}\xspace}
\def\MRad {\ensuremath{\rm \,MRad}\xspace}
\def\ci {\ensuremath{\rm \,Ci}\xspace}
\def\mci {\ensuremath{\rm \,mCi}\xspace}

\def\sx    {\ensuremath{\sigma_x}\xspace}    
\def\sy    {\ensuremath{\sigma_y}\xspace}   
\def\sz    {\ensuremath{\sigma_z}\xspace}    

\newcommand{\stat}{\ensuremath{\mathrm{(stat)}}\xspace}
\newcommand{\syst}{\ensuremath{\mathrm{(syst)}}\xspace}


\def\order{{\ensuremath{\cal O}}\xspace}
\newcommand{\chisq}{\ensuremath{\chi^2}\xspace}

\def\deriv {\ensuremath{\mathrm{d}}}

\def\gsim{{~\raise.15em\hbox{$>$}\kern-.85em
          \lower.35em\hbox{$\sim$}~}\xspace}
\def\lsim{{~\raise.15em\hbox{$<$}\kern-.85em
          \lower.35em\hbox{$\sim$}~}\xspace}

\newcommand{\mean}[1]{\ensuremath{\left\langle #1 \right\rangle}} 
\newcommand{\abs}[1]{\ensuremath{\left\|#1\right\|}} 
\newcommand{\Real}{\ensuremath{\mathcal{R}e}\xspace}
\newcommand{\Imag}{\ensuremath{\mathcal{I}m}\xspace}

\def\PDF {PDF\xspace}

\def\Ebeam {\ensuremath{E_{\mbox{\tiny BEAM}}}\xspace}
\def\sqs   {\ensuremath{\protect\sqrt{s}}\xspace}

\def\ptot       {\mbox{$p$}\xspace}
\def\pt         {\mbox{$p_{\rm T}$}\xspace}
\def\et         {\mbox{$E_{\rm T}$}\xspace}
\def\dpp        {\ensuremath{\mathrm{d}\hspace{-0.1em}p/p}\xspace}

\newcommand{\dedx}{\ensuremath{\mathrm{d}\hspace{-0.1em}E/\mathrm{d}x}\xspace}


\def\dllkpi     {\ensuremath{\mathrm{DLL}_{\kaon\pion}}\xspace}
\def\dllppi     {\ensuremath{\mathrm{DLL}_{\proton\pion}}\xspace}
\def\dllepi     {\ensuremath{\mathrm{DLL}_{\electron\pion}}\xspace}
\def\dllmupi    {\ensuremath{\mathrm{DLL}_{\mmu\pi}}\xspace}

\def\mphi       {\mbox{$\phi$}\xspace}
\def\mtheta     {\mbox{$\theta$}\xspace}
\def\ctheta     {\mbox{$\cos\theta$}\xspace}
\def\stheta     {\mbox{$\sin\theta$}\xspace}
\def\ttheta     {\mbox{$\tan\theta$}\xspace}

\def\degrees{\ensuremath{^{\circ}}\xspace}
\def\krad {\ensuremath{\rm \,krad}\xspace}
\def\mrad{\ensuremath{\rm \,mrad}\xspace}
\def\rad{\ensuremath{\rm \,rad}\xspace}

\def\betastar {\ensuremath{\beta^*}}
\newcommand{\lum} {\ensuremath{\mathcal{L}}\xspace}
\newcommand{\intlum}[1]{\ensuremath{\int\lum=#1\xspace}}  


\def\evtgen     {\mbox{\textsc{EvtGen}}\xspace}
\def\pythia     {\mbox{\textsc{Pythia}}\xspace}
\def\fluka      {\mbox{\textsc{Fluka}}\xspace}
\def\tosca      {\mbox{\textsc{Tosca}}\xspace}
\def\ansys      {\mbox{\textsc{Ansys}}\xspace}
\def\spice      {\mbox{\textsc{Spice}}\xspace}
\def\garfield   {\mbox{\textsc{Garfield}}\xspace}
\def\geant      {\mbox{\textsc{Geant3}}\xspace}
\def\hepmc      {\mbox{\textsc{HepMC}}\xspace}
\def\gauss      {\mbox{\textsc{Gauss}}\xspace}
\def\gaudi      {\mbox{\textsc{Gaudi}}\xspace}
\def\boole      {\mbox{\textsc{Boole}}\xspace}
\def\brunel     {\mbox{\textsc{Brunel}}\xspace}
\def\davinci    {\mbox{\textsc{DaVinci}}\xspace}
\def\erasmus    {\mbox{\textsc{Erasmus}}\xspace}
\def\moore      {\mbox{\textsc{Moore}}\xspace}
\def\ganga      {\mbox{\textsc{Ganga}}\xspace}
\def\dirac      {\mbox{\textsc{Dirac}}\xspace}
\def\root       {\mbox{\textsc{Root}}\xspace}
\def\roofit     {\mbox{\textsc{RooFit}}\xspace}
\def\pyroot     {\mbox{\textsc{PyRoot}}\xspace}
\def\photos     {\mbox{\textsc{Photos}}\xspace}

\def\cpp        {\mbox{\textsc{C\raisebox{0.1em}{{\footnotesize{++}}}}}\xspace}
\def\python     {\mbox{\textsc{Python}}\xspace}
\def\ruby       {\mbox{\textsc{Ruby}}\xspace}
\def\fortran    {\mbox{\textsc{Fortran}}\xspace}
\def\svn        {\mbox{\textsc{SVN}}\xspace}

\def\kbytes     {\ensuremath{{\rm \,kbytes}}\xspace}
\def\kbsps      {\ensuremath{{\rm \,kbytes/s}}\xspace}
\def\kbits      {\ensuremath{{\rm \,kbits}}\xspace}
\def\kbsps      {\ensuremath{{\rm \,kbits/s}}\xspace}
\def\mbsps      {\ensuremath{{\rm \,Mbits/s}}\xspace}
\def\mbytes     {\ensuremath{{\rm \,Mbytes}}\xspace}
\def\mbps       {\ensuremath{{\rm \,Mbyte/s}}\xspace}
\def\mbsps      {\ensuremath{{\rm \,Mbytes/s}}\xspace}
\def\gbsps      {\ensuremath{{\rm \,Gbits/s}}\xspace}
\def\gbytes     {\ensuremath{{\rm \,Gbytes}}\xspace}
\def\gbsps      {\ensuremath{{\rm \,Gbytes/s}}\xspace}
\def\tbytes     {\ensuremath{{\rm \,Tbytes}}\xspace}
\def\tbpy       {\ensuremath{{\rm \,Tbytes/yr}}\xspace}

\def\dst        {DST\xspace}


\def\nonn {\ensuremath{\rm {\it{n^+}}\mbox{-}on\mbox{-}{\it{n}}}\xspace}
\def\ponn {\ensuremath{\rm {\it{p^+}}\mbox{-}on\mbox{-}{\it{n}}}\xspace}
\def\nonp {\ensuremath{\rm {\it{n^+}}\mbox{-}on\mbox{-}{\it{p}}}\xspace}
\def\cvd  {CVD\xspace}
\def\mwpc {MWPC\xspace}
\def\gem  {GEM\xspace}

\def\tell1  {TELL1\xspace}
\def\ukl1   {UKL1\xspace}
\def\beetle {Beetle\xspace}
\def\otis   {OTIS\xspace}
\def\croc   {CROC\xspace}
\def\carioca {CARIOCA\xspace}
\def\dialog {DIALOG\xspace}
\def\sync   {SYNC\xspace}
\def\cardiac {CARDIAC\xspace}
\def\gol    {GOL\xspace}
\def\vcsel  {VCSEL\xspace}
\def\ttc    {TTC\xspace}
\def\ttcrx  {TTCrx\xspace}
\def\hpd    {HPD\xspace}
\def\pmt    {PMT\xspace}
\def\specs  {SPECS\xspace}
\def\elmb   {ELMB\xspace}
\def\fpga   {FPGA\xspace}
\def\plc    {PLC\xspace}
\def\rasnik {RASNIK\xspace}
\def\elmb   {ELMB\xspace}
\def\can    {CAN\xspace}
\def\lvds   {LVDS\xspace}
\def\ntc    {NTC\xspace}
\def\adc    {ADC\xspace}
\def\led    {LED\xspace}
\def\ccd    {CCD\xspace}
\def\hv     {HV\xspace}
\def\lv     {LV\xspace}
\def\pvss   {PVSS\xspace}
\def\cmos   {CMOS\xspace}
\def\fifo   {FIFO\xspace}
\def\ccpc   {CCPC\xspace}

\def\cfourften     {\ensuremath{\rm C_4 F_{10}}\xspace}
\def\cffour        {\ensuremath{\rm CF_4}\xspace}
\def\cotwo         {\ensuremath{\rm CO_2}\xspace} 
\def\csixffouteen  {\ensuremath{\rm C_6 F_{14}}\xspace} 
\def\mgftwo     {\ensuremath{\rm Mg F_2}\xspace} 
\def\siotwo     {\ensuremath{\rm SiO_2}\xspace} 

\newcommand{\eg}{\mbox{\itshape e.g.}\xspace}
\newcommand{\ie}{\mbox{\itshape i.e.}}
\newcommand{\etal}{{\slshape et al.\/}\xspace}
\newcommand{\etc}{\mbox{\itshape etc.}\xspace}
\newcommand{\cf}{\mbox{\itshape cf.}\xspace}
\newcommand{\ffp}{\mbox{\itshape ff.}\xspace}
 
\widetext


\title{Measurements of  {\boldmath$\Bd\to\jpsi\piz$} and other \boldmath{\CP} violating modes at \belle}
\author{Bilas Pal \\ Brookhaven National Laboratory, Upton, NY\\ (On behalf of the Belle Collaboration) }


\begin{abstract}
We report the recent measurements of  $\Bd\to\jpsi\piz$ and other \CP violation modes based on the data collected by the Belle experiment at the KEKB collider.  The \CP asymmetry parameters for the decay $\Bd\to\jpsi\piz$ have previously been measured by \babar and \belle experiments, but the results of mixing induced \CP asymmetry $[S=-\eta_f\sin(2\phi_1)]$ were not in good agreement with each other. Furthermore, the \babar result lies outside the physically allowed region. Previous \belle measurements were based on $535\times10^6$ $\B\Bbar$  pairs. We updated the measurements using the final Belle data set of $772\times10^6$ $\B\Bbar$  pairs.  The \CP asymmetry parameters from a charmless $\Bd\to\KS\piz\piz$ decay using final 
\belle data set and  measurement of $\cos(2\phi_1)$ in $\Bd\to\D^{(*)}h^0$ using joint \babar and \belle analysis are also discussed. 
\end{abstract}

\maketitle

\section{Introduction}
In the standard model (SM) of electroweak interaction,  charge-parity (\CP) violation arises from an  irreducible complex phase
in the Cabibbo-Kobayashi-Maskawa (CKM) quark-mixing matrix~\cite{ckm}.  The \belle and \babar experiments have established \CP violating effects  in the $\B$ meson system. Both experiments use their measurements of the mixing-induced \CP violation in $b\to \ccbar s$ transitions to precisely determine the parameter $\sin (2\phi_1)$, where $\phi_1$ is defined as $\arg[-V_{cd}V^*_{cb}/V_{td}V^*_{tb}]$, with $V_{ij}$ is the CKM matrix element of quarks $i,~j$. In this proceeding an overview  of  recent measurements of the CKM angles  $\phi_1$  is presented. 
\section{Branching fraction and \boldmath{\CP} asymmetries of \boldmath{$\Bd\to\jpsi\piz$}}
\begin{figure}[htb]
\centering
\includegraphics[height=0.85in,width=0.244\textwidth]{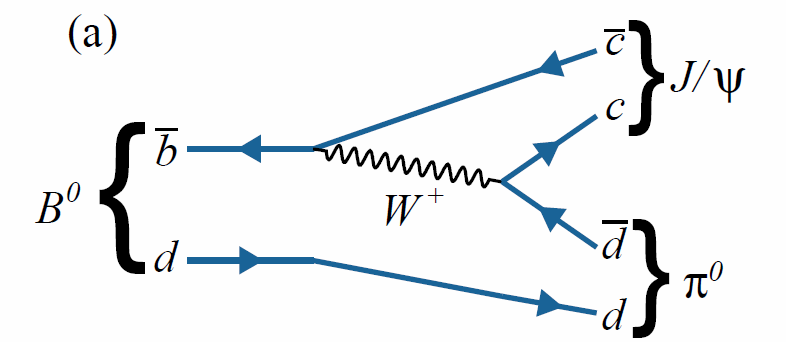}%
\includegraphics[height=0.85in,width=0.244\textwidth]{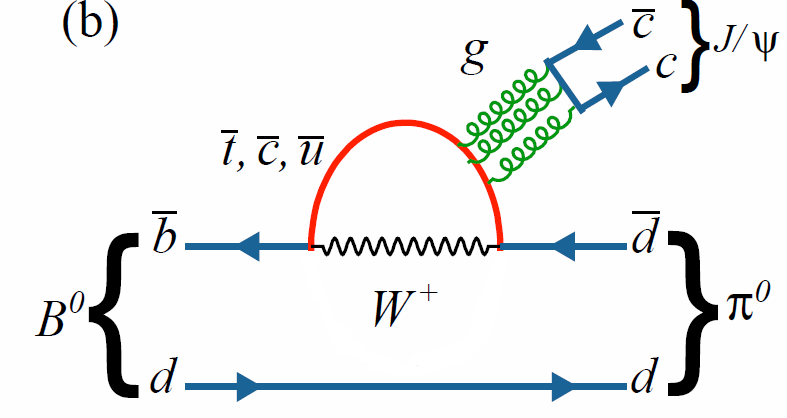}
\caption{\small (a) Tree and (b) penguin amplitudes for the decay $\Bz\to\jpsi\piz$.}
\label{fig:Feynman}
\end{figure}

\tighten
At  the quark level, the decay $\Bz\to\jpsi\piz$ proceeds via $b\to\ccbar d$ 
``tree'' and ``penguin'' amplitudes, as shown in Fig.~\ref{fig:Feynman}. Both 
amplitudes are suppressed in the Standard Model (the first one is color- and Cabibbo-suppressed), 
and thus the branching fraction
is small. The tree-level amplitude has the same weak phase as that of the 
$b\to\ccbar s$ amplitude governing, $e.g.$, $B^0\to \jpsi \KS$ decays,
while the penguin amplitude has a different weak phase. The former dominates mixing-induced \CP violation, while the addition of the latter gives rise to 
direct \CP violation.

In the process $\FourS\to\Bz\Bzb$, one of the two $B$ mesons can decay 
into a \CP eigenstate $f^{}_{\CP}$ at time $t_{\CP}$, while the other can decay 
into a flavor-specific state $f_{\rm tag}$ at time $t_{\rm tag}$.  The decay 
time evolution for the $B\to f^{}_{\CP}$  is~\cite{Carter:1980tk}
\begin{eqnarray}
\label{eq:one}
\mathcal{P}(\Delta t,q) & = & \frac{e^{-|\Delta t|/\tau_{\Bz}}}{4\tau_{\Bz}}\times \\ 
 & & \biggl( 1 + q\bigl[ \mathcal{S}\sin(\Delta m^{}_d\Delta t) 
+ \mathcal{A}\cos(\Delta m^{}_d\Delta t)  \bigr] \biggr),\nonumber
\end{eqnarray}
where 
$\Delta t=t_{\CP}-t_{\rm tag}$ is the difference in proper decay times between the 
two $B$ mesons; $q=+1~(-1)$ for  signal $\Bzb ~(\Bz)$ decays; 
$\Delta m^{}_d$ is the mass difference between the two mass eigenstates of 
the $\Bz$-$\Bzb$ system; and $\tau^{}_{B^0}$ is the $B^0$ lifetime.
The parameters $\mathcal{S}$ and $\mathcal{A}$ are \CP-violating and characterize 
mixing-induced and direct \CP violation, respectively. In the absence of the 
penguin amplitude, $\mathcal{A}=0$ and $\mathcal{S} = -\eta_{f}\sin(2\phi_1)$, where $\eta_{f}$ is
the \CP eigenvalue of the final state (for $\jpsi\piz$ final state, $\eta_f=+1$).
However, this amplitude
 and any new physics (NP) process having a different weak phase  will shift 
$\mathcal{S}$ and $\mathcal{A}$ from these values. Thus, measuring these parameters
provides a way to search for NP. The values of $\mathcal{S}$ and $\mathcal{A}$ 
measured in $\Bz\to\jpsi\piz$ decays can also be used to constrain the small penguin 
contribution to $\Bz\to\jpsi\KS$ decays~\cite{Ciuchini:2005mg, Faller:2008zc, 
Jung:2012mp, DeBruyn:2014oga, Ligeti:2015yma, Frings:2015eva}. This small 
contribution is important as the decay $\Bz\to\jpsi\KS$ provides the 
most precise determination of $\phi^{}_1$.

The parameter $\mathcal{S}$ for $\Bz\to\jpsi\piz$ has previously been measured 
by Belle~\cite{Lee:2007wd} and BaBar~\cite{Aubert:2008bs}, but the results are 
not in good agreement. The BaBar result lies outside the physically allowed 
region, but the uncertainties are large. The previous result from Belle
was based on $535\times 10^6$ $\BBbar$ pairs~\cite{Lee:2007wd}. Here we 
update that measurement using the final Belle data set of $772\times 10^6$ $\BBbar$ 
pairs~\cite{Pal:2018olx}. We also update the $\Bz\to\jpsi\piz$ branching fraction, for which our previous 
measurement used only $32\times 10^6$ $\BBbar$ pairs~\cite{Abe:2002rc}. 
In addition to more data, the analysis presented here also uses improved tracking 
and photon reconstruction.

The \FourS is produced with a Lorentz boost of $\beta\gamma=0.425$ along the $+z$ axis, 
which is defined as anti-parallel to the $\ep$ beam direction.  Since the $\Bz$ and 
$\Bzb$ mesons are approximately at rest in the \FourS center-of-mass (CM) system, 
$\Delta t$ is determined from the displacement in $z$ between the two $B$ decay 
vertices: $\Delta t \approx \Delta z/c\beta\gamma$. The vertex position for the $\Bz\to\jpsi\piz$ decay is reconstructed using lepton tracks from $\jpsi$ decays.
We perform a vertex fit with a constraint to the interaction point (IP) profile. 
A vertex position for $f_{\rm tag}$ is obtained using tracks that are not 
assigned to the  $B^0\to\jpsi\piz$ candidate, plus the IP constraint.
This constraint allows for reconstruction of an $f_{\rm tag}$ vertex even 
in cases when only one track candidate satisfies the requirement on SVD hits. 
We tag (identify) the flavor of the accompanying $B$ meson using inclusive properties 
of particles not associated with the signal $\Bz\to\jpsi\piz$ decay. The 
algorithm for flavor tagging is described in Ref.~\cite{Kakuno:2004cf}. 
we determine the wrong-tag fractions $\omega^{}_l$ ($l=1,7$) and their differences 
$\Delta\omega^{}_l$ between $\Bz$ and $\Bzb$ decays from 
a control sample of self-tagged semileptonic and hadronic $b\to c$ decays~\cite{Abe:2004mz, Adachi:2012et}.

\begin{figure}[h!t!p!]
\begin{center}
    \includegraphics[width=0.45\textwidth]{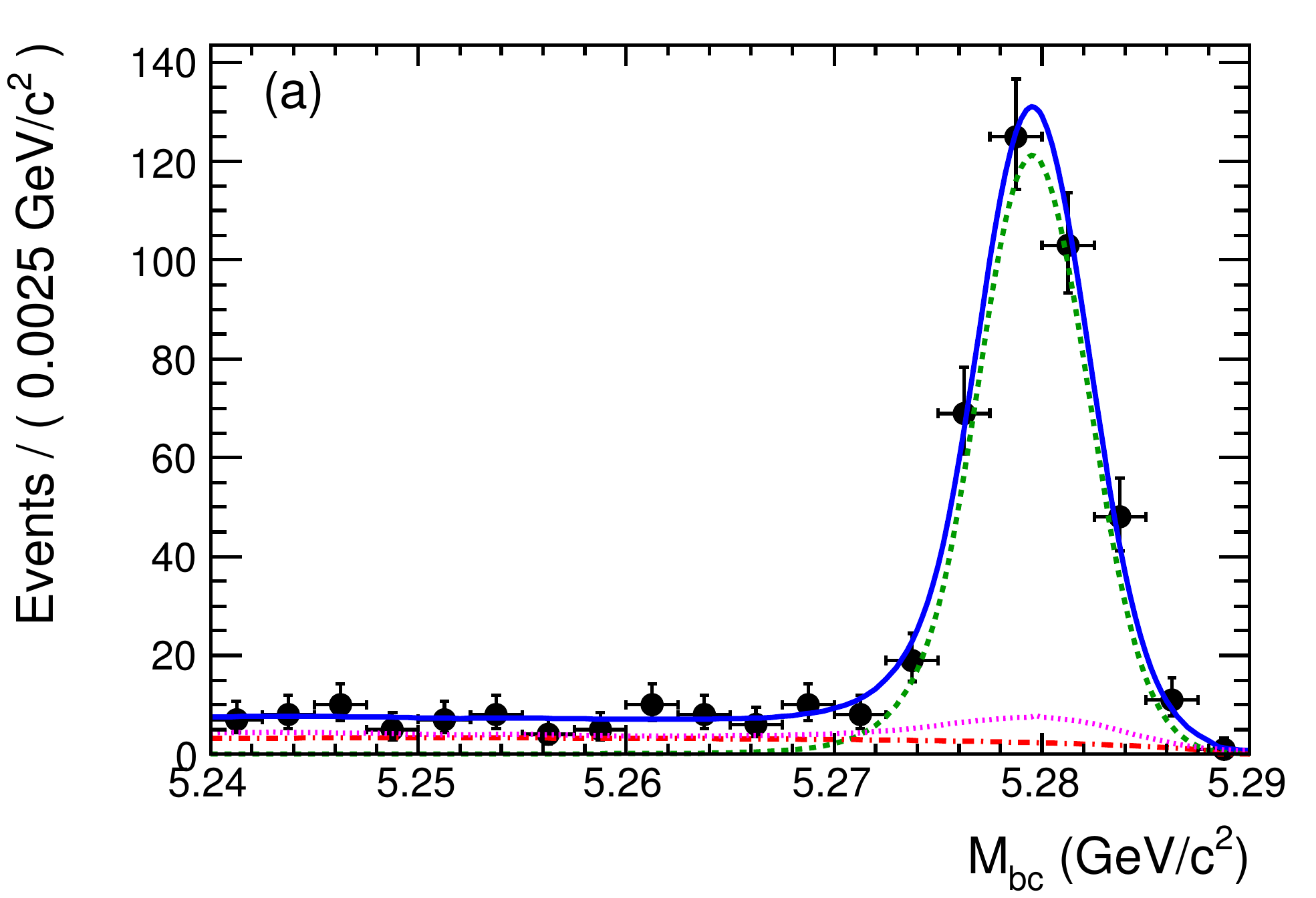}
    \includegraphics[width=0.45\textwidth]{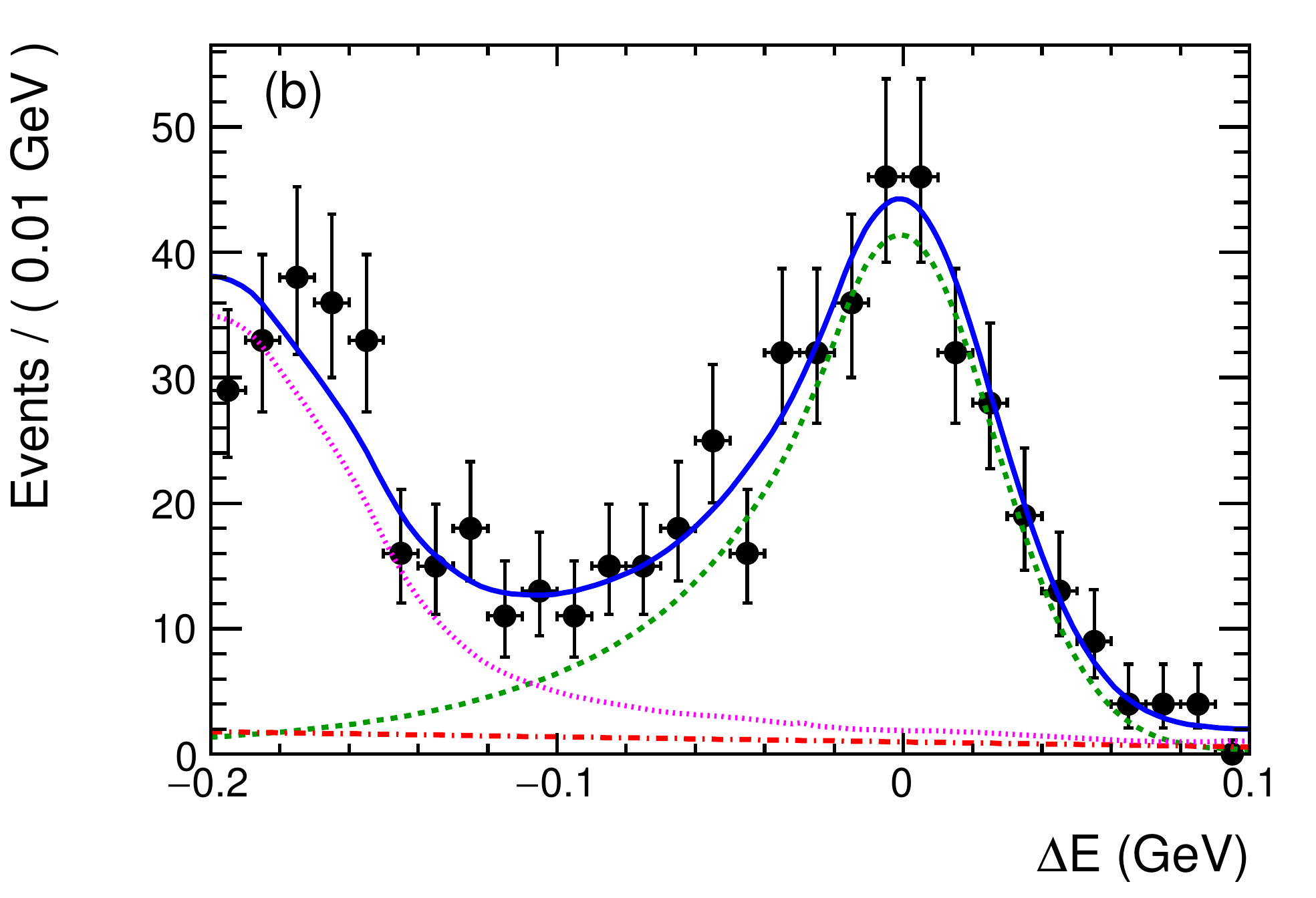}
\end{center}\label{fig:svda_mbc_de_full}
\vskip -0.65cm
\caption{\small Projections of the two-dimensional fit:  
(a) $M_{\rm bc}$ in the $\Delta E$ signal region, and 
(b) $\Delta E$ in the $M_{\rm bc}$ signal region. The points are data, 
the (green) dashed curves show the signal, the (red) dot-dashed curves show the 
$q\bar{q}$ background, the (magenta) dotted curves show the $\BBbar$ background, 
and the (blue) solid curves show the total PDF.}
\end{figure}

Using a fit to the kinematic variables beam-energy constrained mass $M_{\rm bc}=\sqrt{E^2_{\rm beam}-p^2_B}$ and energy difference $\Delta E = E_{\rm beam}-E_B$, where $E_{\rm beam}$ is the beam energy and $p_B$ ($E_B$) is the reconstructed $\B$ meson momentum (energy) in the CM system, we extract $330.2\pm 22.1$ signal events, as shown in Fig.~\ref{fig:svda_mbc_de_full}.
The remaining background is small and dominated by $\BBbar$ events in which 
one of the $B$ mesons decays into a final state containing a $\jpsi$. We divide
this background into three categories: 
(a) $\Bz\to\jpsi\KS$, (b) $\Bz\to\jpsi\KL$, and (c) $B\to\jpsi X$ other than 
$\Bz\to\jpsi\Kz$. 
The rest of the
background  comes from continuum $\qqbar~(q=u,d,s,c)$ events. 

\begin{figure}[h!t!p!]
\begin{center}
    \includegraphics[width=0.45\textwidth]{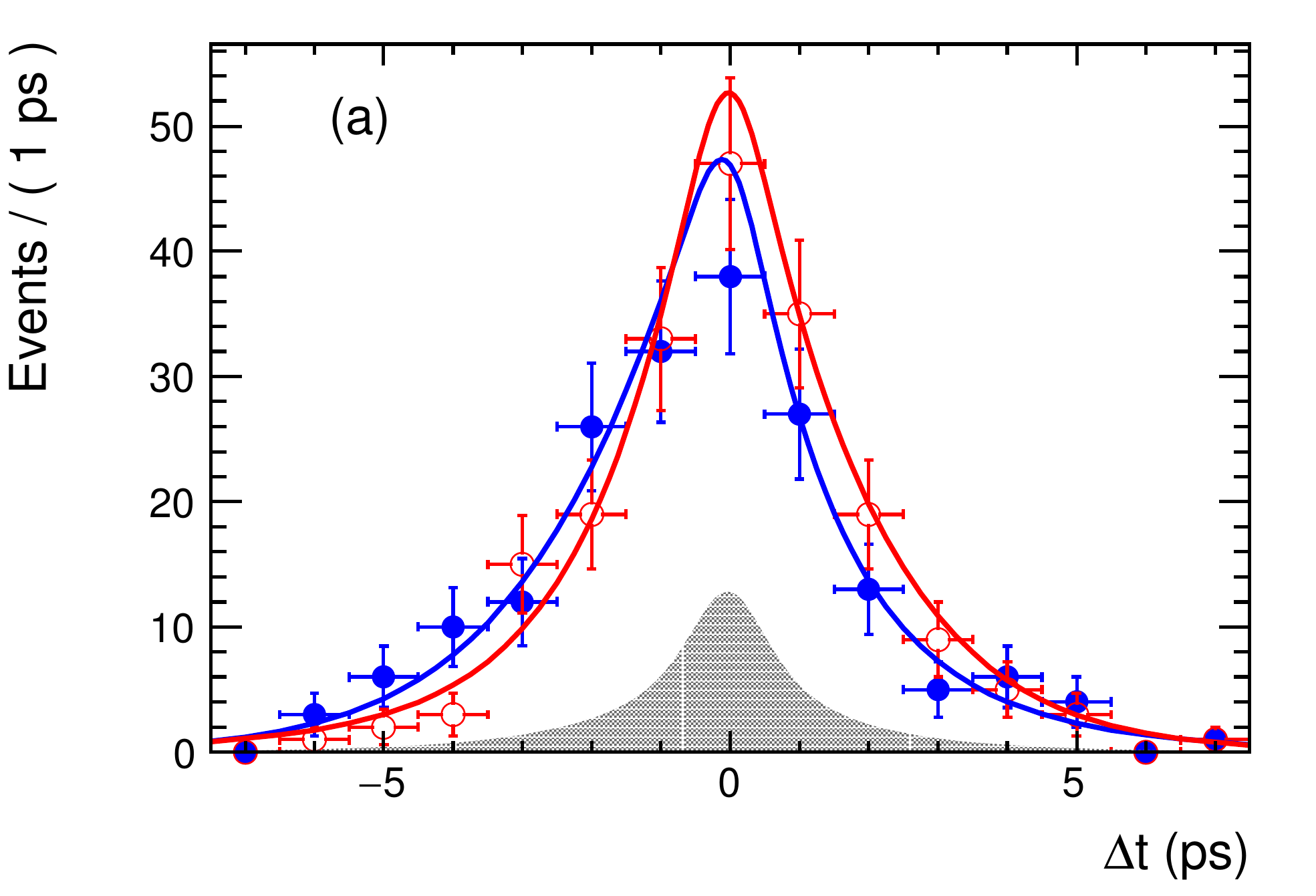}
        \includegraphics[width=0.45\textwidth]{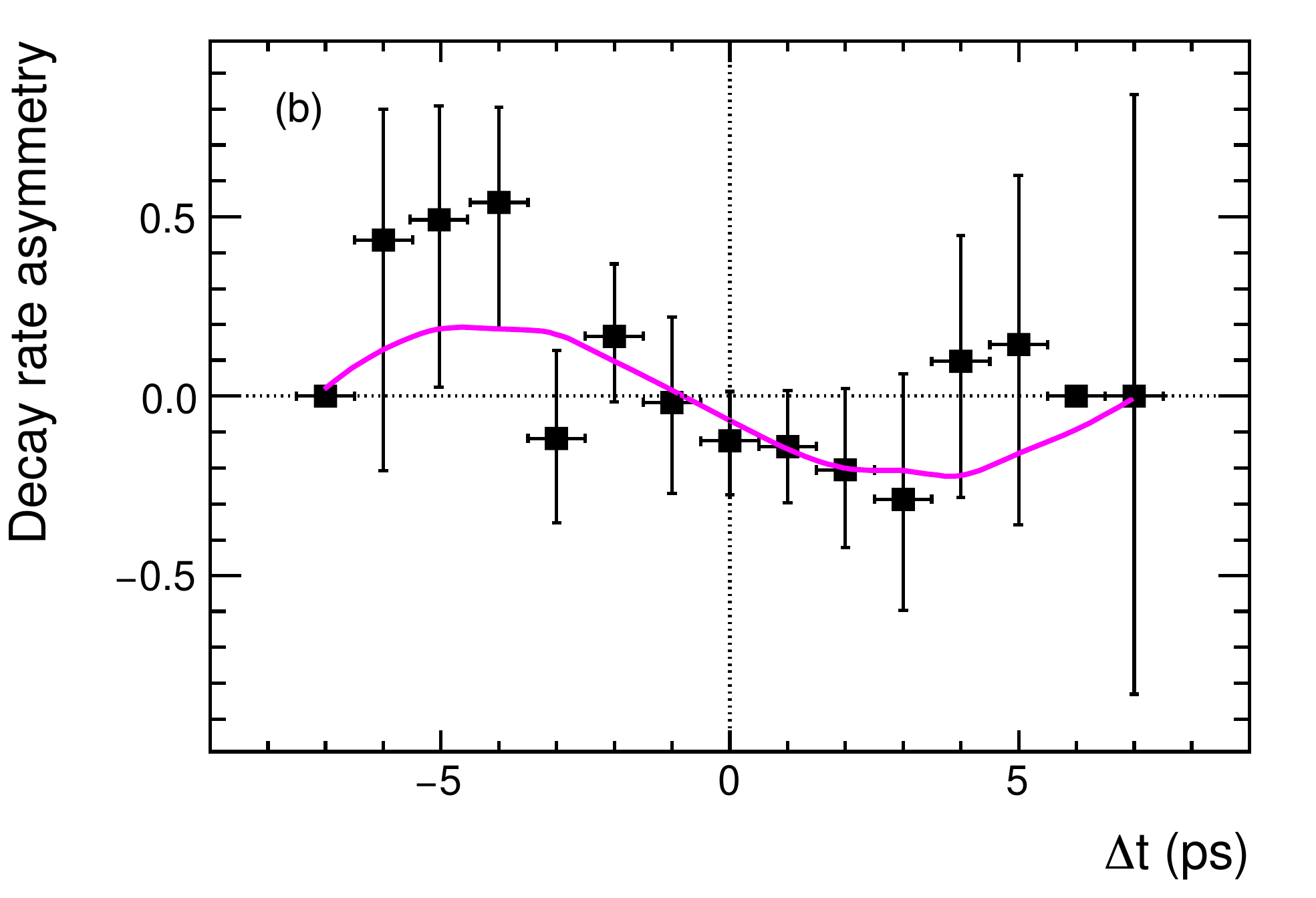}%
\end{center}\label{fig:svda_dt_full}
\vskip -0.65cm
\caption{\small (a) Distributions of  $\Delta t$. The (blue) solid and (red) open 
points represent the $q=+1$ and $q=-1$  events, respectively, and the solid curves 
show the corresponding fit projections. The gray shaded region represents the sum of all
backgrounds.  (b) Time-dependent asymmetry ${\cal A}^{}_{CP}$ (see text). }
\end{figure}

We determine $\mathcal{S}$ and $\mathcal{A}$ by performing an unbinned maximum 
likelihood fit to the $\Delta t$ distribution  of candidate events in the signal 
region. The PDF for the signal component, 
$\mathcal{P}_{\rm sig}(\Delta t; \mathcal{S}, \mathcal{A}, q, \omega_l, \Delta \omega_l)$, 
is given by Eq.~(\ref{eq:one}) with the parameters $\tau_{\Bz}$ and $\Delta m_d$ 
fixed to the world-average values~\cite{Amhis:2014hma}. We modify this expression 
to take into account the effect of incorrect flavor assignment, which is 
parametrized by $\omega_l$ and $\Delta \omega_l$. This PDF is then convolved 
with the decay-time resolution function $R_{\rm sig}(\Delta t)$. The resolution 
function is itself a convolution of four components: 
the detector resolutions for $z^{}_{\jpsi\piz}$ and $z^{}_{\rm tag}$; 
the shift of the $z_{\rm tag}$ vertex position due to secondary tracks from 
charmed particle decays; and the kinematic approximation that the $\B$ mesons 
are at rest in the CM frame~\cite{Adachi:2012et}. The PDFs for $\Bz\to\jpsi\KS$ 
and $\Bz\to\jpsi\KL$ backgrounds are the same as $\mathcal{P}_{\rm sig}$
but with \CP parameters $\mathcal{A}$ and $\mathcal{S}$ fixed to the recent 
\belle results~\cite{Adachi:2012et}. The PDF for $\B\to\jpsi X$ background 
is taken to have the same form as $\mathcal{P}_{\rm sig}$ but with $\mathcal{A}$ 
and $\mathcal{S}$ set to zero, and with an effective lifetime $\tau^{}_{\rm eff}$ 
determined from MC simulation. The PDF for continuum background is taken to
be the sum of two Gaussian functions whose parameters are obtained by fitting events 
in the sideband region $5.20 \gevcc < M_{\rm bc} < 5.26 \gevcc$ and 
$0.10 \gev < \Delta E < 0.50 \gev$.  
Figure~\ref{fig:svda_dt_full} shows the fitted $\Delta t$ distribution and the 
time-dependent decay rate asymmetry $\mathcal{A}_{\CP}$, where 
$\mathcal{A}_{\CP}=
\bigl( Y^{(q=+1)}_{\rm sig} - Y^{(q=-1)}_{\rm sig}\bigr)/
\bigl( Y^{(q=+1)}_{\rm sig} + Y^{(q=-1)}_{\rm sig}\bigr)$, where $Y^{(q=\pm1)}_{\rm sig}$ is the signal yield with $q=\pm1$.

We measure
\begin{eqnarray*}
\mathcal{B} & = & (1.62 \pm 0.11 \pm0.07)\times10^{-5} \\
\mathcal{S} & = & -0.59 \pm 0.19 \pm 0.03  \\
\mathcal{A} & = & -0.15 \pm 0.14\,^{+0.04}_{-0.03}\,,
\end{eqnarray*}
where the first uncertainty is statistical and the second is systematic. 
The measured value for the branching fraction is the most precise value to date and supersedes
the previous measurement~\cite{Abe:2002rc}. It 
is consistent with measurements made by other experiments~\cite{Aubert:2008bs,  Avery:2000yh}. The measured \CP asymmetries 
are consistent with, and supersede, our previous results~\cite{Lee:2007wd}. 
The direct \CP asymmetry $\mathcal{A}$ is consistent with zero.
The mixing-induced \CP asymmetry $\mathcal{S}$ differs from zero ($i.e.$, no \CP violation) 
by 3.0 standard deviations, and it differs from the BaBar result~\cite{Aubert:2008bs}
(which is outside the physical region) 
by 3.2 standard deviations. The value  is consistent 
with the value of $\sin 2\phi_1$ measured using $b\to\ccbar s$ decays~\cite{PDG}. These results
indicate that the penguin and any NP contribution to $\Bz\to\jpsi\piz$ 
are small.
\section{$\sin(2\phi_1^{\rm eff})$ measurement from a \boldmath{\CP}-even state of \boldmath{$\Bd\to\KS\piz\piz$}}
The three-body charmless hadronic $\Bd$ decays to a \CP-even final state $\KS\piz\piz$ mainly proceed via a $b\to\ddbar s$ ``penguin" transition.
Measurements of $\sin(2\phi_1^{\rm eff})$ from such decays are generally sensitive to new physics effects, since the new particles in several extensions of the SM may appear as virtual contributions in the loop.  
We note that there is a $b\to\uubar s$ tree  amplitude also contribute to this decay and can shift $\phi_1^{\rm eff}$ from $\phi_1$. However, this amplitude is doubly Cabibbo-suppressed, and thus the resulting shift is small~\cite{Cheng:2007cz}.
Previously, \babar collaboration studied this decay and measured $\sin(2\phi_1^{\rm eff}) = -0.72\pm 0.71 \pm 0.08$~\cite{Aubert:2007ub}; the sign is opposite to the expectation from the SM, although the statistical uncertainty is large. Here we present the first such measurement using the final Belle data sample~\cite{Yusa:2018hmz}, which is 3.4 times larger than that of the \babar.

\begin{figure}[h!t!p!]
\begin{center}
    \includegraphics[width=0.45\textwidth]{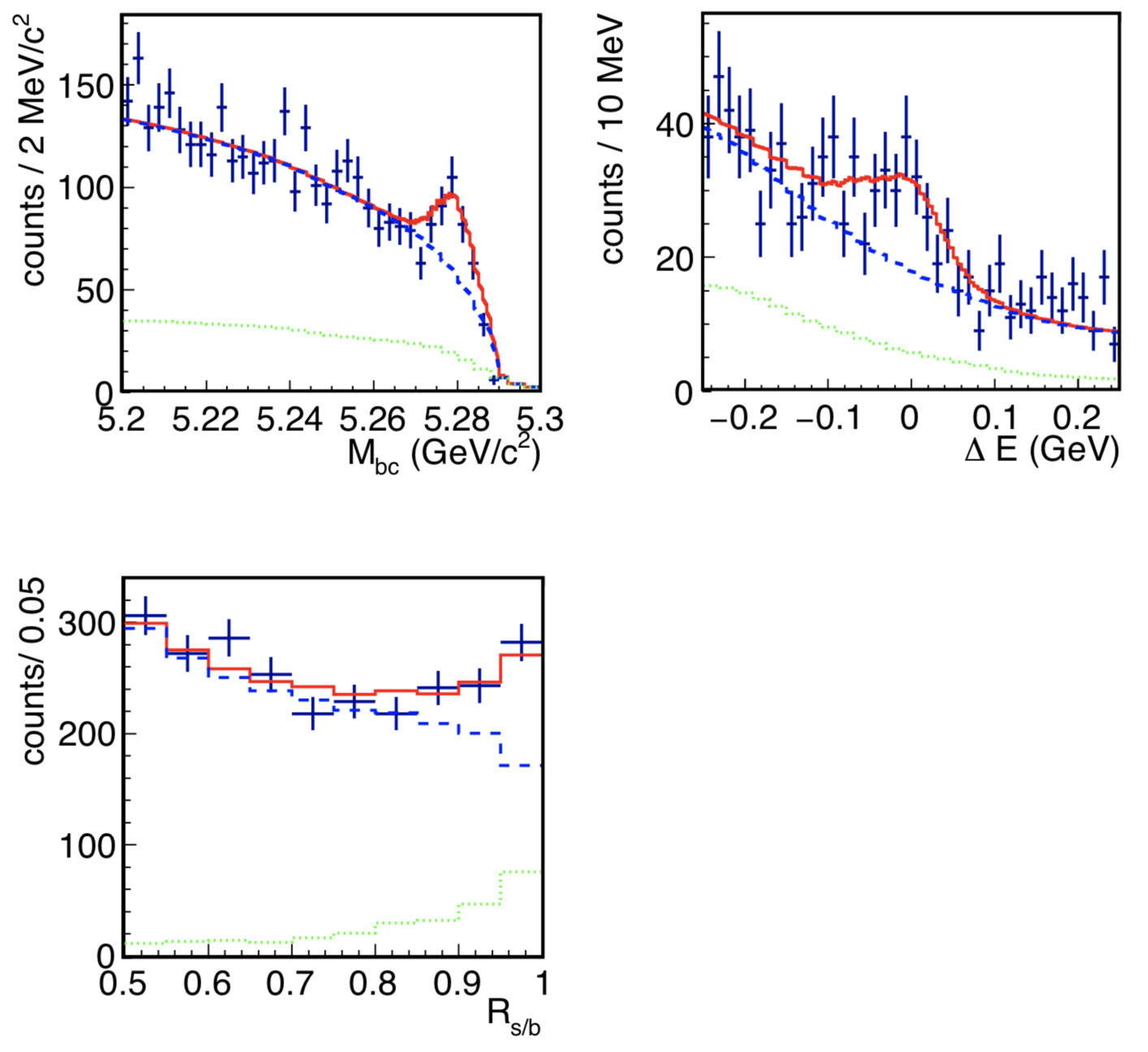}
\end{center}\label{fig:4}
\vskip -0.65cm
\caption{\small (a) Distributions of kinematic and continuum suppression variables (points with uncertainties) using signal-enhanced selection $M_{\rm bc}>5.27$ \gevcc, $-0.15~ \gev <\Delta E <0.10~\gev$, and $R_{s/b}>0.9$ except for the variable displayed. The fit result is illustrated by the (red) solid  curve, while the total and $\B\Bbar$ background are shown by (blue) dashed and  (green) dotted  curves, respectively.}
\end{figure}

\begin{figure}[h!t!p!]
\begin{center}
    \includegraphics[width=0.45\textwidth]{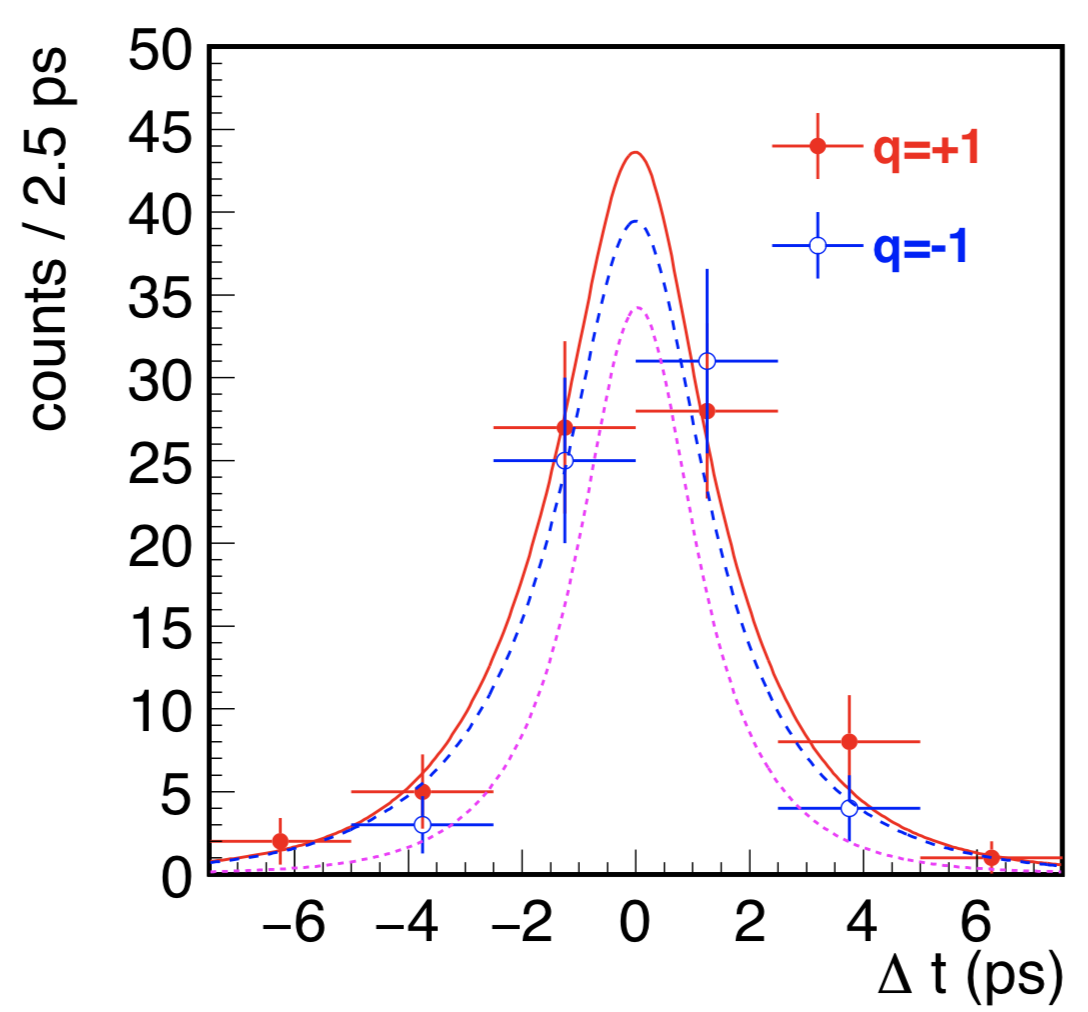}
\end{center}\label{fig:5}
\vskip -0.65cm
\caption{\small $\Delta t$ distribution shown by data points with uncertainties, and the fit results with curves: (red) filled circles with error bars along with a (red) solid-line fit curve corresponds to $q=+1$, while (blue) open circles with error bars along with a (blue) dashed-line fit curve corresponds to $q=-1$. The background contribution is illustrated by (magenta) dotted-line. Events with good flavor tagging quality are shown only.}
\end{figure}

Since no charged tracks from the decay point of the $\Bd$ decays into the \CP eigenstate of $\KS\piz\piz$, the vertex is determined from the direction of $\KS$ and the constraint from IP. From a fit to the variables of kinematic  and continuum suppression, we extract $146.7\pm23.6$ signal events. The projections of the fit are shown in Fig.~\ref{fig:4}.
The dominant source of background originates from continuum events. To suppress this background, which tend to be jet-like from spherical $\B\Bbar$ events,  a likelihood ratio is calculated using the so-called event shape variables. The remaining background originates from $\B\Bbar$ events. From a fit to the $\Delta t$ (see Fig.~\ref{fig:5}), we obtain
\begin{eqnarray*}
\sin(2\phi_1^{\rm eff}) & = & 0.92\,\,^{+0.27}_{-0.31} \pm 0.11 \\
\mathcal{A} & = & 0.28 \pm 0.21\pm0.04\,,
\end{eqnarray*}
where the first uncertainty is statistical and the second is systematic. 
The value of  $\mathcal{S}$ is consistent with the value measure from decays mediated by $b\to\ccbar s$ transition~\cite{PDG}. 
The value of $\mathcal{A}$ is consistent with zero, $i.e.$, no direct \CP violation,
as expected in the SM. 
\section{\boldmath$\phi_1$ measurement in $\Bd\to D^{(*)}h^0$ decays with \babar + \belle data}
By performing time-dependent \CP violation analyses of the ``gold plated" decay mode $\Bd\to\jpsi\KS$ and other decays mediated by $b\to\ccbar s$ transitions, 
$\sin(2\phi_1)$ is measured with a good accuracy~\cite{PDG}.
However, inferring the \CP-violating weak phase $\phi_1$ from the measurement of $\sin(2\phi_1)$ leads to a two-fold trigonometric ambiguity between $2\phi_1$ and $\pi-2\phi_1$ (a four-fold ambiguity in $\phi_1$), and therefore to an ambiguity in the determination of the apex of the CKM Unitarity Triangle. 
This ambiguity can be resolved by measuring $\cos(2\phi_1)$, which is experimentally accessible in $\B$ meson decays that involve multi-body final states. 
However, no previous single measurement has been sensitive enough 
to establish the sign of $\cos(2\phi_1)$, to resolve the ambiguity without further assumptions.

The decays studied here are $\Bd\to D^{(*)}h^0$ with $D\to\KS\pip\pim$, where $h^0$ denotes a light, unflavored, and neutral hadron $(h^0\in\{\piz,\eta,\omega\}$)~\cite{Adachi:2018itz}.
These decays provide an experimentally elegant and powerful way to access $\cos(2\phi_1)$~\cite{Bondar:2005gk}.
To provide a more precise measurement of $\cos(2\phi_1)$, a joint analysis using the combined full dataset of \babar and \belle experiments is performed. 
The $D\to\KS\pip\pim$ decay exhibits complex interference structures that receive resonant and non-resonant contributions to the three-body final state from a rich variety of intermediate \CP eigenstates and quasi-flavor-specific decays. Knowledge of the variations on the relative strong phase as a function of the three-body Dalitz plot phase space enables measurements of both $\sin(2\phi_1)$ and $\cos(2\phi_1)$ from the time evolution of the $\Bd\to [\KS\pip\pim]^{(*)}_{D}h^0$ multi-body final state. 
The Dalitz plot amplitude analysis uses a high-statistics $\epem\to\ccbar$ charm data set collected by \belle that provides about 1.2 million  $D\to\KS\pip\pim$ decays at high purity. The obtained model parameters are  fixed in the subsequent analysis of the $\B$ meson decay.

\begin{figure*}[h!t!p!]
\begin{center}
    \includegraphics[width=0.99\textwidth]{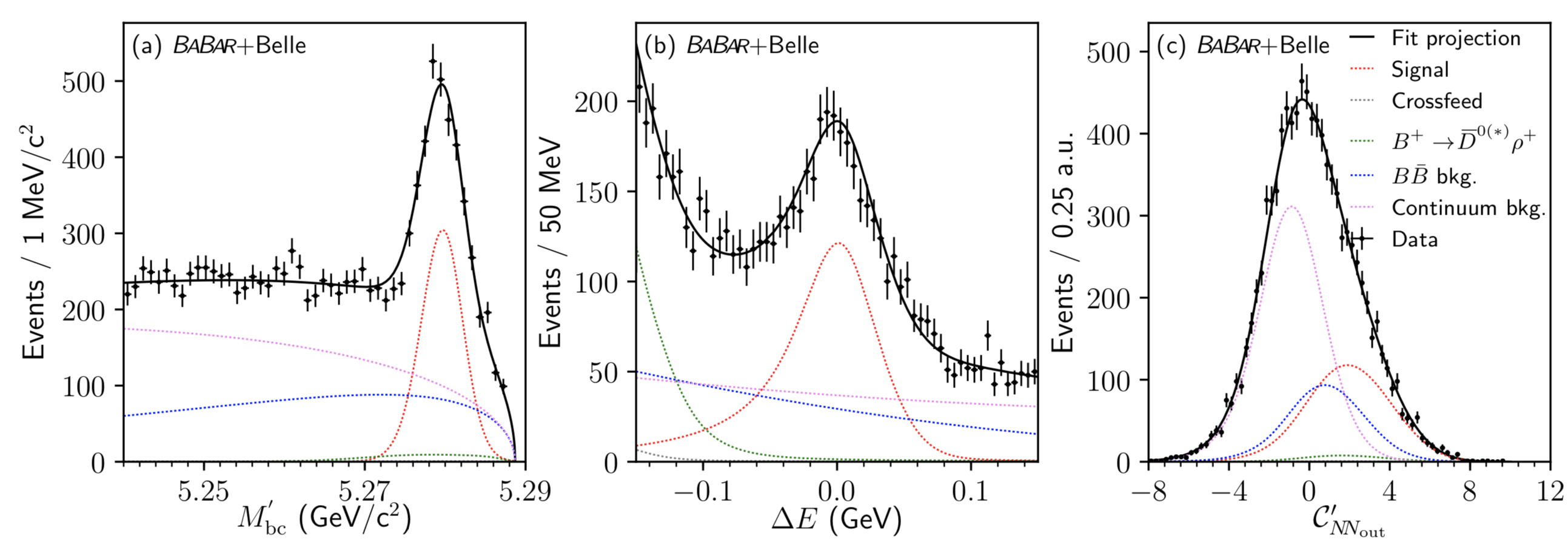}
\end{center}\label{fig:6}
\vskip -0.65cm
\caption{\small Data distributions for (a) $M_{\rm bc}$, (b) $\Delta E$, and (c) $\mathcal{C}_{\rm NN}$ (points with error bars) for the \babar and \belle
data samples combined. The solid black lines represent projections of the total fit function, and the colored dotted lines show the signal and background components of the fit as indicated in the legend. In plotting the  distributions, each of the other two observables are required to satisfy $M_{\rm bc}>5.272~\gevcc$, $|\Delta E|< 100$ MeV, or  $0<\mathcal{C}_{\rm NN}<8$ to select signal-enhanced regions.}
\end{figure*}

In total, a $\Bd\to D^{(*)} h^0$ signal yield of $1129\pm48$ events in the \babar data sample
and $1557\pm 56$ events in the \belle data sample is obtained.  The yields are determined by three-dimensional unbinned maximum 
likelihood fit to the distributions of the observables $M_{\rm bc}$, $\Delta E$, and neural network output  $\mathcal{C}_{\rm NN}$ (neural network is used to suppress the background originates from continuum events). The fit projections are shown in Fig.~\ref{fig:6}.

\begin{figure}[h!t!p!]
\begin{center}
    \includegraphics[width=0.5\textwidth]{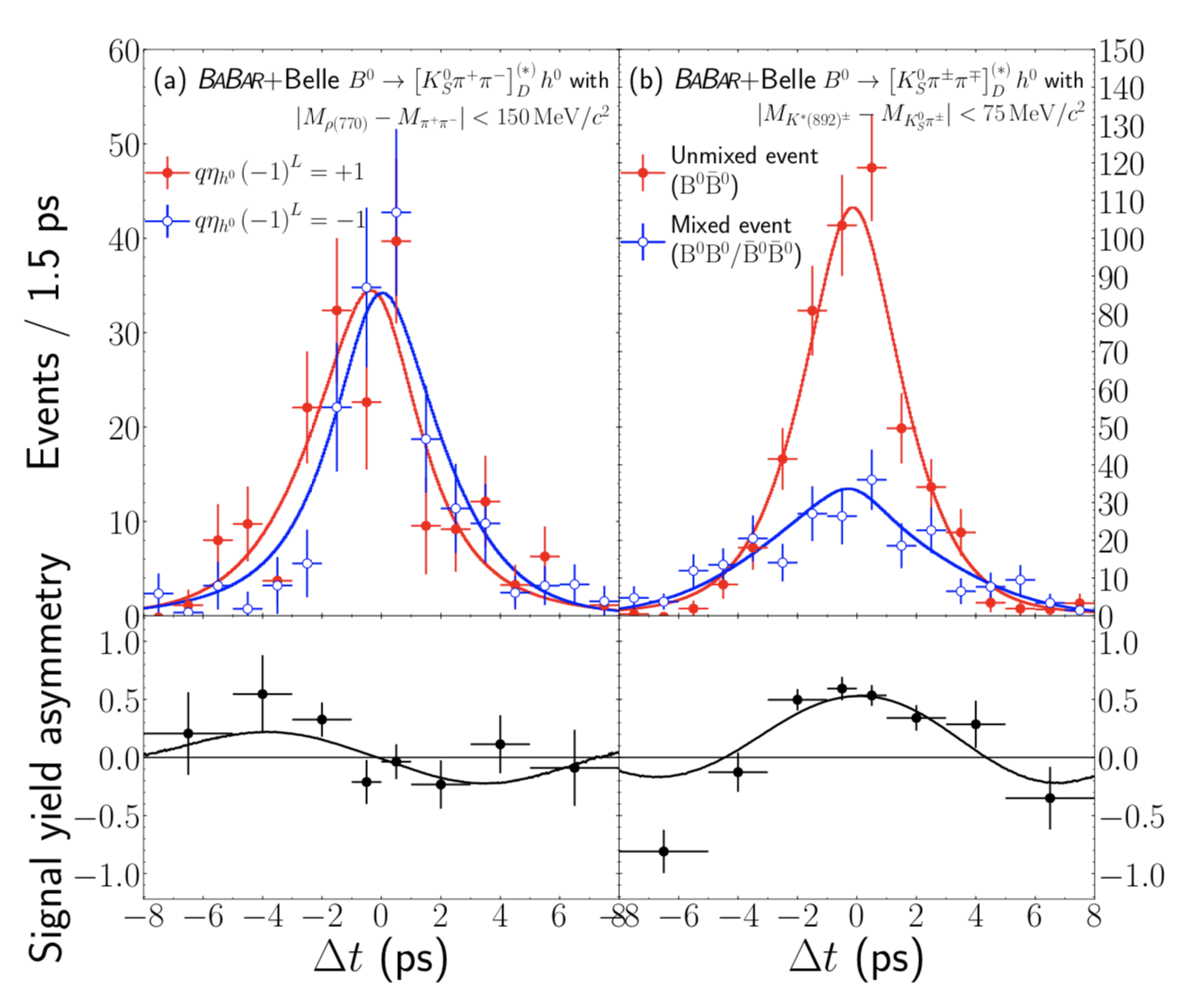}
\end{center}\label{fig:7}
\vskip -0.65cm
\caption{\small Distributions of the proper time interval (data points with error bars) and the corresponding asymmetries for $\Bd\to D^{(*)}h^0$ candidates associated with high quality flavor tags for two different regions of the $D\to\KS\pip\pim$ phase space and for the \babar and \belle data samples combined. The background has been subtracted using the sPlot technique, with weights obtained from the fit presented in Fig.~\ref{fig:6}.}
\end{figure}

From a fit to the $\Delta t$ distribution of the selected events, we obtain
\begin{eqnarray*}
\sin(2\phi_1) & = & 0.80 \pm 0.14 \pm 0.06 \pm 0.03  \\
\cos(2\phi_1) & = & 0.91\pm 0.22 \pm 0.09 \pm 0.07,
\end{eqnarray*}
where the first uncertainty is statistical, the second is systematic and the third is due to the $D\to\KS\pip\pim$ decay amplitude model. 
A direct measurement of the angle $\phi_1$ results in
\begin{eqnarray*}
\phi_1 & = & (22.5\pm 4.4 \pm 1.2 \pm 0.6)^{\circ}.
\end{eqnarray*}
Distribution of proper  time intervals and its asymmetries are shown in Fig.~\ref{fig:7}.
The results provide a first evidence for  $\cos(2\phi_1) >0$ at the level of 3.7 standard deviations, and  excludes the trigonometric multifold solution 
$\pi/2-\phi_1=(68.1\pm0.7)^{\circ}$ at the level of 7.3 standard deviations. 
Thus the measurements directly rules out the unfavored solution of the Unitarity Triangle. CP violation is observed at the level of 5.1 standard deviations
\section{Summary}
Recent measurements of $\Bd\to\jpsi\piz$ and other \CP violating modes ($\Bd\to\KS\piz\piz$ and $\Bd\to D^{(*)}h^0$) using the final dataset of \belle experiments ($\Bd\to D^{(*)}h^0$ analysis is performed using combined \babar and \belle dataset) are presented. 
The results of $\sin(2\phi_1)$ from all the three decay modes are consistent with the $\sin(2\phi_1)$ measured in $b\to\ccbar s$ processes. 
The measurement of $\cos(2\phi_1)$ in $\Bd\to D^{(*)}h^0$ resolves the ambiguity in the determination of the apex of the CKM Unitarity Triangle.
\\
\\
\textbf{Acknowledgement}:
The author thanks the workshop organizers for hosting a fruitful and stimulating workshop
and providing excellent hospitality. This research is supported by the U.S. Department of Energy.


\begin{thebibliography}{99}
\bibitem{ckm}
  N.~Cabibbo,
  Phys.\ Rev.\ Lett.\  {\bf 10}, 531 (1963);
  M.~Kobayashi and T.~Maskawa,
  Prog.\ Theor.\ Phys.\  {\bf 49}, 652 (1973).

\bibitem{Carter:1980tk} 
  A.~B.~Carter and A.~I.~Sanda,
  {\it \CP violation in $B$ meson decays},
  Phys.\ Rev.\ D {\bf 23}, 1567 (1981);
  I.~I.~Y.~Bigi and A.~I.~Sanda,
  {\it Notes on the observability of \CP violations in $B$ decays},
  Nucl.\ Phys.\ B {\bf 193}, 85 (1981).

\bibitem{Faller:2008zc} 
  S.~Faller, M.~Jung, R.~Fleischer and T.~Mannel,
  {\it The golden modes $\Bz \to\jpsi K_{S,L}$ in the era of precision flavor physics},
  Phys.\ Rev.\ D {\bf 79}, 014030 (2009)
  [arXiv:0809.0842 [hep-ph]].
  
\bibitem{Jung:2012mp} 
  M.~Jung,
  {\it Determining weak phases from $B\to J/\psi P$ decays},
  Phys.\ Rev.\ D {\bf 86}, 053008 (2012)
  [arXiv:1206.2050 [hep-ph]].
  
\bibitem{DeBruyn:2014oga} 
  K.~De Bruyn and R.~Fleischer,
  {\it A roadmap to control penguin effects in $B^0_d\to J/\psi K_{\rm S}^0$ and $B^0_s\to J/\psi \phi$},
  JHEP {\bf 1503}, 145 (2015)
  [arXiv:1412.6834 [hep-ph]].
  
\bibitem{Ligeti:2015yma} 
  Z.~Ligeti and D.~J.~Robinson,
 {\it Towards more precise determinations of the quark mixing phase $\beta$},
  Phys.\ Rev.\ Lett.\  {\bf 115}, 251801 (2015)
  [arXiv:1507.06671 [hep-ph]].

\bibitem{Ciuchini:2005mg} 
  M.~Ciuchini, M.~Pierini and L.~Silvestrini,
  {\it The Effect of penguins in the $B_d \to \jpsi K^0$ \CP asymmetry},
  Phys.\ Rev.\ Lett.\  {\bf 95}, 221804 (2005)
  [hep-ph/0507290].

\bibitem{Frings:2015eva} 
  P.~Frings, U.~Nierste and M.~Wiebusch,
  {\it Penguin contributions to \CP phases in $B_{d,s}$ decays to charmonium},
  Phys.\ Rev.\ Lett.\  {\bf 115}, 061802 (2015)
  [arXiv:1503.00859 [hep-ph]].

\bibitem{Lee:2007wd} 
  S.~E.~Lee {\it et al.} (Belle Collaboration),
  {\it Improved measurement of time-dependent \CP violation in $\Bz \to \jpsi\piz$ decays},
  Phys.\ Rev.\ D {\bf 77}, 071101 (2008)
  [arXiv:0708.0304 [hep-ex]].
  
  
\bibitem{Aubert:2008bs} 
  B.~Aubert {\it et al.} (BaBar Collaboration),
 {\it Evidence for \CP violation in $B^0 \to J/\psi \pi^0$ decays},
  Phys.\ Rev.\ Lett.\  {\bf 101}, 021801 (2008)
  [arXiv:0804.0896 [hep-ex]].
  
  
\bibitem{Pal:2018olx} 
  B.~Pal {\it et al.} [Belle Collaboration],
  arXiv:1810.01356 [hep-ex].
  
\bibitem{Abe:2002rc} 
  K.~Abe {\it et al.} (Belle Collaboration),
  {\it Measurement of branching fractions and charge asymmetries for two-body $B$ meson decays with charmonium},
  Phys.\ Rev.\ D {\bf 67}, 032003 (2003)
  [hep-ex/0211047].


\bibitem{Kakuno:2004cf} 
  H.~Kakuno {\it et al.} (Belle Collaboration),
  {\it Neutral $B$ flavor tagging for the measurement of mixing induced \CP violation at Belle},
  Nucl.\ Instrum.\ Meth.\ A {\bf 533}, 516 (2004)
  [hep-ex/0403022].


\bibitem{Abe:2004mz} 
  K.~Abe {\it et al.} (Belle Collaboration),
 {\it Improved measurement of CP-violation parameters $\sin 2\phi_1$ and $|\lambda|$, $B$ meson lifetimes, and $\Bz - \Bzb$ mixing parameter $\Delta m_d$},
  Phys.\ Rev.\ D {\bf 71}, 072003 (2005)
  Erratum: [Phys.\ Rev.\ D {\bf 71}, 079903 (2005)]
  [hep-ex/0408111].
  
\bibitem{Adachi:2012et} 
  I.~Adachi {\it et al.} (Belle Collaboration),
  {\it Precise measurement of the \CP violation parameter $\sin 2\phi_1$ in $B^0\to(c\bar c)K^0$ decays},
  Phys.\ Rev.\ Lett.\  {\bf 108}, 171802 (2012)
  [arXiv:1201.4643 [hep-ex]].

\bibitem{Amhis:2014hma} 
  Y.~Amhis {\it et al.} (Heavy Flavor Averaging Group Collaboration),
  {\it Averages of $b$-hadron, $c$-hadron, and $\tau$-lepton properties as of summer 2016},
  Eur.\ Phys.\ J.\ C {\bf 77}, 895 (2017)
  [arXiv:1612.07233 [hep-ex]].

\bibitem{Avery:2000yh} 
  P.~Avery {\it et al.} (CLEO Collaboration),
 {\it Study of exclusive two-body $B^0$ meson decays to charmonium},
  Phys.\ Rev.\ D {\bf 62}, 051101 (2000)
  [hep-ex/0004032].
  
  \bibitem{PDG}
 M. Tanabashi {\it et al.}  (Particle Data Group),
 {\it Review of Particle Physics},
  Phys. Rev.  D {\bf 98}, 030001 (2018). 
  
\bibitem{Cheng:2007cz} 
  H.~Y.~Cheng,
  {\it Theoretical issues with $\Delta \mathcal{S}$ in three-body $\B$ decays},
  hep-ph/0702252 [hep-ph].
  
\bibitem{Aubert:2007ub} 
  B.~Aubert {\it et al.} (BaBar Collaboration),
  {\it Measurement of \CP asymmetry in $B^0 \to K_s \pi^0 \pi^0$ decays},
  Phys.\ Rev.\ D {\bf 76}, 071101 (2007)
  [hep-ex/0702010].
  
\bibitem{Yusa:2018hmz} 
  Y.~Yusa {\it et al.} (Belle Collaboration),
 {\it Measurement of time-dependent \CP violation in $B^0 \to K^0_S \pi^0 \pi^0$ decays},
  [arXiv:1810.03336 [hep-ex]].
  
\bibitem{Adachi:2018itz} 
  I.~Adachi {\it et al.} (BaBar and Belle Collaborations),
 {\it First evidence for $\cos 2\beta>0$ and resolution of the CKM Unitarity Triangle ambiguity by a time-dependent Dalitz plot analysis of $B^{0} \to D^{(*)} h^{0}$ with $D \to K_{S}^{0} \pi^{+} \pi^{-}$ decays},
  arXiv:1804.06152 [hep-ex].
  
\bibitem{Bondar:2005gk} 
  A.~Bondar, T.~Gershon and P.~Krokovny,
  {\it A Method to measure $\phi_1$ using $\Bdb \to D h^0$ with multibody $D$ decay},
  Phys.\ Lett.\ B {\bf 624}, 1 (2005)
  [hep-ph/0503174].
  
\end{thebibliography}
\end{document}